\documentclass[
superscriptaddress,
amsmath,amssymb,
preprintnumbers,
twocolumn,
pra,
aps,
floatfix,longbibliography
]{revtex4-2}
\usepackage{placeins}
\usepackage{flushend}
\usepackage{booktabs}
\usepackage{multirow}
\usepackage{siunitx}
\usepackage{graphicx}
\usepackage{dcolumn}
\usepackage{bm}
\usepackage{chngcntr} 
\usepackage[pdftex,colorlinks=true, linkcolor = blue, citecolor=blue,urlcolor=blue, bookmarksnumbered=true, bookmarksopen=true]{hyperref}
\usepackage[mathlines]{lineno}
\usepackage{physics}
\usepackage{tikz}
\usepackage{quantikz}
\usepackage{algpseudocode}
\usepackage{algorithm}
\usepackage{booktabs}
\usepackage{adjustbox}
\usepackage{soul}
\usepackage{multirow}
\usepackage{comment}
\usepackage{longtable}
\usepackage{booktabs}
\usepackage{siunitx}

\usetikzlibrary{positioning,fit,backgrounds}

\definecolor{wireblue}{RGB}{35,92,115}
\definecolor{givensblue}{RGB}{190,225,238}
\definecolor{hoplavender}{RGB}{232,205,232}
\definecolor{onsitegray}{RGB}{215,215,215}
\definecolor{slaterbg}{RGB}{246,246,246}
\definecolor{evolbg}{RGB}{231,244,226}
\definecolor{ctrlgray}{RGB}{115,115,120}

\definecolor{elegantBlue}{RGB}{31, 78, 121}
\definecolor{elegantRed}{RGB}{128, 54, 45}
\definecolor{softGray}{RGB}{248, 248, 248}
\definecolor{bondGray}{RGB}{60, 60, 60}

\newcounter{myalgorithm}

\begin{document}

\title{Quantum Algorithms to Determine Spin-Resolved Exchange-Correlation Potential for Strongly Correlated Materials}
\author{H. Arslan Hashim}
\email{Corresponding author, email: HafizArslan.Hashim@ucf.edu}
\affiliation{Department of Physics, University of Central Florida, Orlando, FL 32816, USA}
\author{Volodymyr M. Turkowski}
\affiliation{Department of Physics, University of Central Florida, Orlando, FL 32816, USA}
\author{Eduardo R. Mucciolo}
\email{Corresponding author, email: Eduardo.Mucciolo@ucf.edu}
\affiliation{Department of Physics, University of Central Florida, Orlando, FL 32816, USA}

\date{\today}

\begin{abstract}
Accurate exchange-correlation (XC) potentials are essential for density functional theory, yet reliable approximations remain challenging for strongly correlated systems. In this work, we present a quantum algorithmic framework to determine spin-resolved XC potentials using a variational quantum eigensolver. Using the Hubbard model as a prototypical strongly correlated lattice system, we prepare ground states in fixed spin sectors through a Hamiltonian variational ansatz initialized from the non-interacting $(U=0)$ Slater-determinant ground state. From the resulting many-body ground states, we extract the XC energy and compute the corresponding spin-resolved XC potentials via finite differences. The accuracy of the approach is benchmarked against exact diagonalization for one- and two-dimensional Hubbard systems of various lattice sizes. We demonstrate that the variational ansatz reproduces the ground-state energies and densities with high fidelity, enabling accurate construction of both magnetic and non-magnetic XC potentials. We analyzed the dependence of the XC potentials on the interaction strength, charge, spin densities, and magnetization.
We also present an empirical complexity scaling relation for the computational cost of the method at a fixed fidelity. These results illustrate how quantum computing can be used to construct spin-resolved XC functionals for correlated lattice models, providing a potential pathway for improving density functional approximations in strongly correlated materials.
\end{abstract}
\maketitle
\section{Introduction}
\label{sec:introduction}
Density functional theory (DFT) is one of the most widely used theoretical frameworks for studying electronic and structural properties of materials. By reformulating the many-body Schrödinger equation in terms of the electron density, DFT reduces the interacting problem to an effective single-particle system described by the Kohn-Sham equations~\cite{Hohenberg1964,Kohn1965}. In this formulation, all many-body correlation effects are contained in the exchange-correlation (XC) functional. While practical approximations such as the local density approximation (LDA)~\cite{Perdew1981} and generalized gradient approximation (GGA)~\cite{Perdew1996} have achieved remarkable success for weakly correlated systems, they often fail to accurately describe strongly correlated materials such as transition-metal oxides, rate-earth and actinide compounds, including Mott insulators~\cite{Anisimov1991}. In these systems, localized electronic interactions and magnetic correlations play a central role and require more accurate XC potentials.

Spin-dependent density functional theory (SDFT) extends the formalism to systems with spin polarization~\cite{barthSpinDFT1972}, enabling the description of magnetic materials and partially spin-filled electronic structures. In this case, the XC functional depends explicitly on both spin densities, leading to spin-resolved XC potentials that determine the effective Kohn-Sham Hamiltonian. Constructing accurate spin-dependent XC potentials therefore remains an important challenge for strongly correlated systems. For lattice models such as the Hubbard model, these ideas form the basis of lattice density functional theory (LDFT), where the site occupations play the role of the density variables~\cite{Lima2003,Capelle2013}.

Quantum computing offers a promising route to address this problem. Variational quantum algorithms such as the variational quantum eigensolver (VQE)~\cite{Peruzzo2014} provide a hybrid quantum-classical approach to compute ground states of interacting many-body Hamiltonians. In VQE, a parameterized quantum circuit prepares a trial state whose energy expectation value is minimized through a classical optimization loop. By employing physically motivated ansätze that capture entanglement and correlation efficiently, VQE can represent strongly correlated states that are difficult to treat using classical approaches~\cite{Stanisic2022}. These features make VQE particularly attractive for studying correlated lattice models and extracting many-body properties relevant to electronic structure.

Several recent works have explored connections between quantum computing and DFT. Early proposals demonstrated how the Hohenberg-Kohn framework can be implemented on quantum devices~\cite{Gaitan2009}. Subsequent developments introduced hybrid quantum-classical embedding schemes and quantum-assisted approaches to derive density functionals and Kohn-Sham potentials~\cite{Bauer2016,Baker2020,Rossmannek2021,Pemmaraju2022,Senjean2023,Ko2023}. More recently, Sheridan \textit{et al.} proposed a hybrid quantum-enhanced density functional theory (QEDFT) framework that uses quantum simulations to improve XC functionals for correlated systems~\cite{Sheridan2024}. These works collectively establish the potential of combining quantum simulations with DFT concepts to improve descriptions of strongly correlated materials.

In this paper, we develop a quantum algorithmic framework for determining spin-resolved XC potentials using VQE. We focus on the fermionic Hubbard model as a prototypical strongly correlated lattice system and employ a Hamiltonian variational ansatz (HVA)~\cite{Wecker2015}. The Hubbard model provides a minimal description of strongly correlated electrons and has long served as a benchmark system for XC approximations in lattice DFT and for comparing many-body computational methods~\cite{Carrascal2015,Essler2005}. Ground states are prepared in fixed particle-number sectors $(N_\uparrow,N_\downarrow)$ using a direct and continuation strategy that gradually increases the interaction strength from the non-interacting limit. From the resulting ground states we compute the XC energy and extract spin-resolved XC potentials through finite differences. The method is benchmarked against exact diagonalization for one- and two-dimensional Hubbard systems of various lattice sizes. Within this range of lattice sizes, we obtain a scaling law for the computational cost of the method as a function of size for a fixed fidelity. The results are encouraging, in the sense that the computational cost of our quantum algorithm scales favorably even for system sizes beyond those accessible via exact diagonalization.

Our results demonstrate that variational quantum simulations can accurately reproduce XC quantities for correlated lattice models. This approach provides a route toward constructing improved spin-dependent XC functionals using quantum simulations, potentially extending density functional methods to regimes where classical approaches become computationally prohibitive.

The remainder of this paper is organized as follows. Section~\ref{sec:dft_vqe_overview} briefly reviews the theoretical background of DFT and the VQE. In Sec.~\ref{sec:model_method} we introduce the Hubbard model and describe the quantum algorithm used to prepare ground states and extract the XC potential. Section~\ref{sec:results} presents numerical results for both spin-unpolarized and spin-polarized systems and analyzes the resulting XC potentials. Finally, Sec.~\ref{sec:conclusion} summarizes the main conclusions and discusses possible future directions.
\section{Background}
\label{sec:dft_vqe_overview}

\subsection{Spin-Polarized Lattice Density Functional Theory}

LDFT provides a mapping between an
interacting many-body problem on a discrete lattice and an auxiliary
non-interacting system that reproduces the same ground-state density~\cite{Gunnarsson1986,Schonhammer1987}.
In the spin-generalized case (SDFT)~\cite{barthSpinDFT1972}, the fundamental variables are the spin-resolved site occupations $n_{i\sigma}$ with $\sigma\in\{\uparrow,\downarrow\}$ and $i$ running over lattice sites. For a lattice with $L$ sites, the Hamiltonian is written as
\begin{equation}
    \hat{H} = \hat T + \hat H_{\mathrm{int}} + \hat H_{\mathrm{ext}},
\end{equation}
where $\hat T$ is the hopping (kinetic energy) operator, $\hat H_{\mathrm{int}}$ contains the
interaction terms, and 
$
    \hat H_{\mathrm{ext}}
    = \sum_{i,\sigma}
      v^{\mathrm{ext}}_{i\sigma}\,\hat n_{i\sigma}.
$
is the spin-dependent external potential.
This potential allows spatial inhomogeneity and spin polarization to be treated on equal footing. The lattice Hohenberg-Kohn theorem establishes a unique mapping between $\{n_{i\sigma}\}$ and the external potential
$\{v^{\mathrm{ext}}_{i\sigma}\}$~\cite{Gunnarsson1986,Schonhammer1987,Coe2015}, where
\begin{equation*}
    n_{i\sigma}
    = \langle \psi_{\mathrm{GS}} | \hat n_{i\sigma} | \psi_{\mathrm{GS}} \rangle.
\end{equation*}
is the ground-state spin density.

Once the density-potential mapping is established, the ground-state energy may be regarded as a functional of the spin densities,
\begin{equation}
    E[\{n_{i\sigma}\}]
    =
    F[\{n_{i\sigma}\}]
    + \sum_{i,\sigma} v^{\mathrm{ext}}_{i\sigma}\, n_{i\sigma},
\end{equation}
where the universal functional
\begin{equation*}
    F[\{n_{i\sigma}\}]
    =
    \langle \psi_{\mathrm{GS}} |
    \hat T + \hat H_{\mathrm{int}}
    | \psi_{\mathrm{GS}} \rangle
\end{equation*}
contains all interaction and kinetic contributions and is independent of the external potential.

Because $F$ is not known explicitly, one introduces the
Kohn-Sham (KS) system~\cite{Kohn1965}, which reproduces the interacting
densities using a non-interacting Hamiltonian. The KS decomposition reads
\begin{align}
E[\{n_{i\sigma}\}]
&=
T_{\mathrm{S}}[\{n_{i\sigma}\}]
+E_{\mathrm{H}}[\{n_{i\sigma}\}]
+E_{\mathrm{XC}}[\{n_{i\sigma}\}]
\nonumber\\
&\quad
+\sum_{i,\sigma}
v^{\mathrm{ext}}_{i\sigma}n_{i\sigma},
\label{eq:energy}
\end{align}
where \(T_\mathrm{S}\) denotes the kinetic-energy functional of the
auxiliary non-interacting Kohn--Sham system that reproduces the interacting
densities. The Hartree functional \(E_H\) represents the classical
mean-field contribution of the electron-electron interaction, while all
remaining many-body effects beyond this mean-field description are
contained in the XC functional \(E_{\mathrm{XC}}\).

Because the lattice density is represented by a finite set of site
occupations, functional derivatives reduce to ordinary partial
derivatives. Variational minimization of the total energy functional yields the effective spin-dependent KS potential,
\begin{equation}
    V^{\mathrm{KS}}_{i\sigma}
    =
    v^{\mathrm{ext}}_{i\sigma}
    + V_{\mathrm{H},i}[\{n_{j}\}]
    + V_{\mathrm{XC},i\sigma}[\{n_{j\sigma'}\}],
    \label{eq:v_kohnsham}
\end{equation}
where the Hartree and XC potentials are defined as
\begin{equation*}
    V_{\mathrm{XC},i\sigma}
    =
    \frac{
    \partial E_{\mathrm{XC}}[\{n_{j\sigma'}\}]
    }{
    \partial n_{i\sigma}
    },
\end{equation*}
\begin{equation*}
    V_{\mathrm{H},i}
    =
    \frac{
    \partial E_{\mathrm{H}}[\{n_{j}\}]
    }{
    \partial n_{i}
    }.
\end{equation*}
The non-interacting kinetic-energy functional \(T_\mathrm{S}\) is evaluated from the auxiliary KS orbitals and is represented by the kinetic operator
\(\hat T\) in the KS Hamiltonian, while the remaining
density-dependent contributions define the effective KS potential. The associated KS Hamiltonian is
\begin{equation}
    \hat H^{\mathrm{KS}}
    = \hat T + \sum_{i,\sigma}
      V^{\mathrm{KS}}_{i\sigma}\,\hat n_{i\sigma},
\end{equation}
with orbitals satisfying
\begin{equation}
\hat h^{\mathrm{KS}}_{\sigma}\,\phi_{j\sigma}
    = \varepsilon_{j\sigma}\,\phi_{j\sigma},
\label{eq:ks_eigenproblem}
\end{equation}
where 
\begin{equation*}
\hat h^{\mathrm{KS}}_{\sigma} = \hat t + \sum_{i} V^{\mathrm{KS}}_{i\sigma} \ket{i}\bra{i},
\end{equation*}
is the single-particle KS Hamiltonian.
The KS spin densities are obtained from the occupied KS orbitals as
$
    n_{i\sigma}^{\mathrm{KS}}
    =
    \sum_{j \in \mathrm{occ}}
    |\phi_{j\sigma}(i)|^2.
$
The total energy may equivalently be expressed in terms of the occupied
KS eigenvalues as
\begin{align}
E[\{n_{i\sigma}\}]
&=
\sum_{\sigma}\sum_{j\in\mathrm{occ}} \varepsilon_{j\sigma} -\sum_{i} V_{\mathrm{H},i}\,n_{i}
-\sum_{i,\sigma} V_{\mathrm{XC},i\sigma}\,n_{i\sigma}
\nonumber\\
&\quad
+ E_{\mathrm{H}}[\{n_{i}\}]
+ E_{\mathrm{XC}}[\{n_{i\sigma}\}].
\label{eq:energy_total}
\end{align}
Here $\varepsilon_{j\sigma}$ are the occupied eigenvalues of the single-particle KS lattice Hamiltonian.

Self-consistency is achieved by iterating the KS cycle and mixing densities
between iterations,
$
    n_{i\sigma}^{(j+1)}
    =
    \alpha\, n_{i\sigma}^{(j)}
    + (1-\alpha)\, n^{\mathrm{KS}}_{i\sigma},
$
until both the density and energy converge. Here, $\alpha\in[0,1]$ is a density-mixing parameter used to stabilize the self-consistent iteration. Once convergence is achieved, the ground-state observables follow from the converged KS orbitals and densities. The complete self-consistent procedure is summarized in
Algorithm~\ref{alg:spin_ldft}.

%----------------------------------------------------------------------
% Algorithm environment
%----------------------------------------------------------------------

\begin{figure}[t]
\centering
\begin{minipage}{\columnwidth}
\rule{\linewidth}{0.4pt}
\refstepcounter{myalgorithm}
\textbf{Algorithm 1: Spin-Polarized LDFT Self-Consistency}
\label{alg:spin_ldft}
\rule{\linewidth}{0.4pt}

\begin{algorithmic}[1]

\State Specify \(v^{\mathrm{ext}}_{i\sigma}\) and initialize
\(n^{(0)}_{i\uparrow},n^{(0)}_{i\downarrow}\).

\Repeat

    \State Construct \(V^{\mathrm{KS}}_{i\sigma}\) from
    Eq.~\eqref{eq:v_kohnsham}.

    \State Solve the KS eigenproblem, Eq.~\eqref{eq:ks_eigenproblem},
    and occupy the lowest-energy orbitals.

    \State Compute \(n_{i\sigma}^{\mathrm{KS}}\) from the occupied KS
    orbitals.

    \State Evaluate the total energy using Eq.~\eqref{eq:energy_total}.

    \State Mix densities:
    \(n_{i\sigma}^{(j+1)}
    =
    \alpha n_{i\sigma}^{(j)}
    +
    (1-\alpha)n_{i\sigma}^{\mathrm{KS}}\).

\Until{density and total energy converge}

\end{algorithmic}

\rule{\linewidth}{0.4pt}
\end{minipage}
\end{figure}

%%%%%%%%%%%%%%%%%%%%%%%%%%%%%%%%%%%%%%%%%%%%%%%%%%%%%%%%%%%%%%%%%%%%%%%%%%%%%
\subsection{Variational Quantum Eigensolver}

The VQE is a hybrid quantum-classical
algorithm designed to approximate low-energy eigenstates of quantum
many-body Hamiltonians using near-term quantum devices
\cite{Peruzzo2014,McClean2016,Tilly2022,Cerezo2021}. It is based on the
Rayleigh-Ritz variational principle: for any normalized trial state
$|\Psi\rangle$, the expectation value
$\langle \Psi|\hat H|\Psi\rangle$ provides an upper bound to the exact
ground-state energy of the Hamiltonian $\hat H$. In VQE, the trial state is
prepared by a parameterized quantum circuit,
\begin{equation}
    |\Psi(\boldsymbol{\theta})\rangle
    =
    \hat U(\boldsymbol{\theta})|\Phi_0\rangle ,
\end{equation}
where $|\Phi_0\rangle$ is a chosen reference state,
$\hat U(\boldsymbol{\theta})$ is a parameterized unitary circuit, and
$\boldsymbol{\theta}$ is a set of classical variational parameters. The
energy functional
\begin{equation}
    E(\boldsymbol{\theta})
    =
    \langle \Psi(\boldsymbol{\theta})|
    \hat H
    |\Psi(\boldsymbol{\theta})\rangle
\end{equation}
is then minimized by a classical optimizer.

A VQE calculation consists of four central components.
The first is the \emph{objective function}, which is usually the energy
expectation value of the target Hamiltonian, although other cost functions
or penalty terms may be added to enforce symmetries or target excited
states \cite{McClean2016,Endo2021,Tilly2022}. The second is the
\emph{parameterized circuit ansatz}, which determines the variational
manifold accessible to the quantum state. The ansatz may be hardware
efficient, chemically or physically motivated, adaptive, or based on
Hamiltonian evolution \cite{Peruzzo2014,McClean2016,Cerezo2021,Tilly2022}.
The third ingredient is the \emph{measurement procedure}: after mapping the
Hamiltonian to a qubit operator, expectation values are estimated by
measuring Pauli terms or groups of commuting Pauli terms. Measurement cost,
statistical fluctuations, and Pauli-grouping strategies are important practical considerations for hardware implementations \cite{Tilly2022,Endo2021}. The fourth
component is the \emph{classical optimizer}, which updates
$\boldsymbol{\theta}$ using the measured objective function. The optimizer
may be gradient-free or gradient-based, and its performance depends on the
landscape of the variational problem, the presence of noise, and possible
trainability issues such as barren plateaus \cite{McClean2016,Cerezo2021}.

For fermionic lattice Hamiltonians, two aspects are especially important:
the fermion-to-qubit encoding and the choice of ansatz. The mapping
determines the Pauli-string structure of the qubit Hamiltonian, while the
ansatz determines how efficiently the relevant many-body correlations can
be represented. For the Hubbard model, Hamiltonian-inspired ansatz are
particularly natural because they preserve particle number and spin
sectors while reflecting the structure of the hopping and interaction
terms.

In the present work, VQE is used as a many-body solver within the
spin-polarized lattice-DFT framework. For each fixed spin sector
$(N_\uparrow,N_\downarrow)$, the optimized VQE state provides the
interacting ground-state energy and observables needed to construct the
XC energy and the corresponding spin-resolved
XC potentials. Since the calculations reported here are
performed using exact noiseless simulation, expectation values are evaluated
directly from the state vector rather than through finite-shot sampling.
Consequently, the method section focuses on the Hubbard-specific implementation of VQE: the fermion-to-qubit encoding, the Hamiltonian-inspired ansatz, the direct and continuation strategy in the interaction strength, and the finite difference extraction of spin-resolved XC potentials.

\section{Method}
\label{sec:model_method}

This section describes the computational workflow used to obtain
spin-resolved XC quantities from VQE ground-state
calculations of the Fermi--Hubbard model. For a given lattice geometry,
the Hubbard Hamiltonian is first mapped to a qubit Hamiltonian using the
Jordan--Wigner transformation (JWT). The resulting qubit Hamiltonian is then
used for all spin sectors $(N_\uparrow, N_\downarrow)$. For each sector,
the workflow consists of: (i) preparing a non-interacting reference state
with the desired spin occupations, (ii) optimizing a finite-(U) HVA, and (iii) extracting
$E_{\rm XC}$ and $V_{\rm XC}$ from the resulting many-body energies
and densities. The individual components of this procedure are described
below.

\subsection{Hubbard model and fixed spin sectors}

We consider the spinful Hubbard model on a lattice with $L=L_x \times L_y$ sites,
\begin{equation}
    \hat H = \hat H_t + \hat H_U ,
    \label{eq:Hubbard}
\end{equation}
where
\begin{equation}
    \hat H_t =
    -t \sum_{\langle i,j\rangle,\sigma}
    \left(
    \hat c^\dagger_{i\sigma}\hat c_{j\sigma}
    +
    \hat c^\dagger_{j\sigma}\hat c_{i\sigma}
    \right)
\end{equation}
is the hopping term with sum over nearest-neighbors, and
\begin{equation}
    \hat H_U =
    U\sum_i \hat n_{i\uparrow}\hat n_{i\downarrow}
\end{equation}
is the onsite interaction. Here
$\hat c^\dagger_{i\sigma}$ and $\hat c_{i\sigma}$ create and annihilate, respectively,
a fermion on site $i$ with spin $\sigma$, and
$\hat n_{i\sigma}=\hat c^\dagger_{i\sigma}\hat c_{i\sigma}$. Where $(t)$ is the nearest-neighbor hopping amplitude and $(U)$ is the onsite interaction strength.

The VQE calculations are performed independently for each spin sector
$(N_\uparrow,N_\downarrow)$. Although the same qubit Hamiltonian is used
for all sectors of a given lattice geometry, the initial state is prepared
with the desired particle numbers and the HVA preserves these quantum
numbers throughout the optimization. This sector-by-sector treatment is
essential because the spin-resolved XC energy is constructed as a function of the spin occupations, while the corresponding
XC potentials are obtained from finite differences
between neighboring spin sectors.

An alternative approach is to project the Hamiltonian explicitly onto a
fixed particle-number sector before mapping it to qubits, which can reduce
the Hilbert-space dimension~\cite{Moll2016} hence the number of qubits. However, depending on the projection and encoding, this can increase
operator nonlocality and the number of Pauli terms in the qubit
Hamiltonian, potentially increasing circuit and measurement cost. We therefore use the
sector-preserving ansatz approach in the main calculations; the
projection-based reduction is discussed in Appendix~\ref{appendix:projection}.
%
%\vspace{-0.3cm}

%%%%%%%%%%%%%%%%%%%%%%%%%%%%%%%%%%%%%%%%%%%%%%%%%%%%%%%%%%%%%%%%%%%%%%%%%%%%
\begin{figure*}[!t]
\centering
\includegraphics[width=0.95\linewidth]{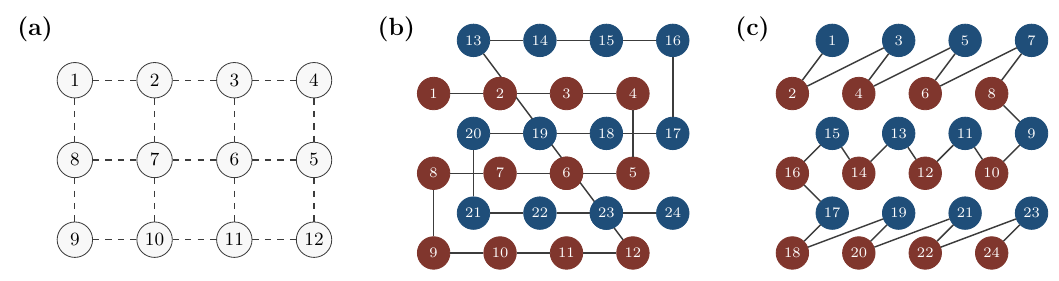}
\caption{
Physical lattice and JW orbital orderings for a
\(3\times4\) spinful Hubbard system. Each lattice site contains two spin
orbitals, so the \(3\times4\) lattice is mapped to a \(2\times3\times4\)
qubit layout corresponding to \(24\) qubits. The numbers in panels
(b) and (c) denote the qubit indices assigned to the fermionic modes.
Red and blue circles represent the two spin sectors.
(a) Physical \(3\times4\) lattice with site labels. The dashed gray bonds
show the nearest-neighbor connectivity of the Hubbard model.
(b) Snake ordering, where the two spin sectors are placed in separate
snake-ordered layers. In this ordering, the two spin orbitals of a given
site are separated between the two layers; for example, site \(1\) maps to
qubits \(1\) and \(13\). This ordering tends to keep same-spin hopping
terms local along the snake path.
(c) Zig-zag ordering, where the spin orbitals are interleaved site by site
along the same physical snake path. In this ordering, the two spin
orbitals of each site are adjacent in the JW ordering; for
example, site \(1\) maps to qubits \(1\) and \(2\). Thus, the zig-zag
ordering localizes onsite interactions, while same-spin hopping terms can
acquire intermediate JW \(Z\)-strings when the corresponding
modes are nonadjacent in the ordering.
}
\label{fig:lattice_ordering}
\end{figure*}
%%%%%%%%%%%%%%%%%%%%%%%%%%%%%%%%%%%%%%%%%%%%%%%%%%%%%%%%%%%%%%%%%%%%%%%%%

\subsection{Fermion-to-qubit encoding}
\label{sec:fermionic}

For a chosen lattice geometry, the fermionic Hubbard Hamiltonian must be
mapped to a qubit Hamiltonian before it can be implemented in a quantum
circuit. Fermion-to-qubit mappings encode fermionic occupation and parity
information into qubit degrees of freedom while preserving the
anticommutation relations of the fermionic creation and annihilation
operators. Several mappings have been developed for this purpose,
including the JWT, parity, Bravyi-Kitaev, and related
low-locality encodings~\cite{Jordan1928,Bravyi2002,Seeley2012,
Tranter2015,Havlicek2017,Setia2019,Steudtner2019}. In this work we use
the JWT, which provides a direct
occupation-number encoding and is widely used in quantum algorithms for
the Fermi-Hubbard model~\cite{Cade2020,Stanisic2022}.

For an \(L\)-site spinful Hubbard system, each lattice site contains two
spin orbitals, \((i,\uparrow)\) and \((i,\downarrow)\). Therefore, the
system contains \(2L\) fermionic modes and is mapped to \(2L\) qubits. We denote the qubit index assigned to the fermionic mode \((i,\sigma)\)
by $q=\pi(i,\sigma)$,
where \(\pi\) specifies the chosen ordering of the spin orbitals.

Under the JWT, the fermionic annihilation and creation
operators on mode \(q\) are mapped to Pauli operators as
\begin{equation}
    \hat c_q =
    \frac{1}{2}
    \left(\prod_{k<q} \hat Z_k\right)
    (\hat X_q-i\hat Y_q)
\end{equation}
and
\begin{equation}
    \hat c_q^\dagger =
    \frac{1}{2}
    \left(\prod_{k<q} \hat Z_k\right)
    (\hat X_q+i\hat Y_q),
\end{equation}
respectively, and the number operator becomes
\begin{equation*}
    \hat n_q
    =
    \hat c_q^\dagger \hat c_q
    =
    \frac{\hat I-\hat Z_q}{2}.
\end{equation*}
The product of the \(\hat Z\) operators stores the fermionic parity of all modes
preceding \(q\) in the chosen ordering and ensures the correct fermionic
anticommutation relations. Consequently, the Pauli structure of the Hamiltonian depends on both the lattice connectivity and the chosen one-dimensional orbital ordering.

The onsite Hubbard interaction is diagonal in the occupation basis. For
site \(i\), let $q_{\uparrow}=\pi(i,\uparrow)$ and $q_{\downarrow}=\pi(i,\downarrow)$. Then,
\begin{equation}
\label{eq:onsite_jw}
    \hat n_{i\uparrow}\hat n_{i\downarrow}
    =
    \frac{1}{4}
    \left(
    \hat I
    -\hat Z_{q_{\uparrow}}
    -\hat Z_{q_{\downarrow}}
    +\hat Z_{q_{\uparrow}}\hat Z_{q_{\downarrow}}
    \right).
\end{equation}
Thus the onsite term maps to a diagonal \(\hat Z \hat Z\)-type operator. Importantly,
it does not contain an intermediate JW parity string. If the two spin
orbitals of a site are far apart in the chosen ordering, the corresponding
two-qubit gate may require additional routing on hardware, but the
operator itself remains a two-qubit diagonal interaction.

In contrast, hopping terms are sensitive to the ordering. For two
same-spin modes $p=\pi(i,\sigma)$ and $q=\pi(j,\sigma)$, the Hermitian hopping operator maps as
\begin{equation}
\label{eq:hopping_jw}
    \hat c_p^\dagger \hat c_q
    +
    \hat c_q^\dagger \hat c_p
    =
    \frac{1}{2}
    \left(
    \hat X_p \hat X_q+ \hat Y_p \hat Y_q
    \right)
    \prod_{k=\min(p,q)+1}^{\max(p,q)-1} \hat Z_k .
\end{equation}
If \(p\) and \(q\) are adjacent in the JW ordering, the intermediate
parity string is absent. If they are not adjacent, the hopping term
contains a string of \(\hat Z\) operators over the modes lying between them.
This ordering dependence is especially important for two-dimensional
Hubbard lattices because any JW mapping embeds the two-dimensional
fermionic lattice into a one-dimensional qubit ordering. Therefore, unlike
a one-dimensional chain, no simple JW ordering can make all nearest-neighbor
hopping terms adjacent simultaneously. The ordering instead determines the
distribution of JW string lengths across horizontal and vertical hopping
terms~\cite{Cade2020,Stanisic2022,Steudtner2019,Hagge2020}.

The orderings illustrated in Fig.~\ref{fig:lattice_ordering} represent
two complementary choices for arranging the fermionic modes in the JW mapping. The snake ordering favors locality of same-spin
hopping terms along the snake path, while the zig-zag ordering localizes
the onsite interaction by placing opposite-spin orbitals of the same site
adjacent in the qubit ordering. Consequently, the choice of ordering
changes the distribution of JW parity-string lengths across
the hopping and interaction terms and therefore affects the structure of
the HVA circuit introduced in the next subsection.

\subsection{Hamiltonian variational ansatz for the Hubbard model}
\label{sec:hva}

Following the JWT described in
Sec.~\ref{sec:fermionic}, we construct the variational circuit using a HVA. The HVA is a problem-inspired
ansatz in which the circuit is built from exponentials of operator groups
appearing in the target Hamiltonian. It was introduced as a short-depth
variational alternative to adiabatic state preparation and was benchmarked
for Hubbard-model systems and small molecules~\cite{Wecker2015}. Related
Hamiltonian-inspired and QAOA-type ans\"atze have since been studied for
fermionic lattice models and near-term Hubbard simulations
\cite{Cade2020,Wiersema2020,Stanisic2022}.

In our implementation, each HVA layer is composed of onsite-interaction
gates and hopping gates. The hopping gates are generated by the
JW-mapped tunneling terms, while the onsite gates are
generated by the diagonal Hubbard interaction. For a hopping term between
two same-spin modes \(p\) and \(q\), the gate has the form
\begin{equation}
    \hat U_{pq}^{(t)}(\theta)
    =
    \exp\left[
    -i\theta
    \left(
    \hat X_p \hat X_q + \hat Y_p \hat Y_q
    \right)
    \prod_{k=\min(p,q)+1}^{\max(p,q)-1} \hat  Z_k
    \right],
    \label{eq:hva_hopping_gate}
\end{equation}
where the sign of the hopping amplitude and the factor of \(1/2\) from the JWT are absorbed into the variational angle \(\theta\). The intermediate JW
\(Z\)-string is determined by the orbital ordering discussed in
Sec.~\ref{sec:fermionic} and illustrated in
Fig.~\ref{fig:lattice_ordering}. If \(p\) and \(q\) are adjacent in the
ordering, the product over \(\hat Z_k\) is absent.

The onsite interaction on site \(i\), acting on the two opposite-spin
modes \(q_{\uparrow}\) and \(q_{\downarrow}\), is implemented as
\begin{equation}
    \hat U_i^{(U)}(\theta_U)
    =
    \exp\left[
    -i\theta_U
    \left(
    -\hat Z_{q_{\uparrow}}
    -\hat Z_{q_{\downarrow}}
    +\hat Z_{q_{\uparrow}} \hat Z_{q_{\downarrow}}
    \right)
    \right],
    \label{eq:hva_onsite_gate}
\end{equation}
up to a global phase from the identity term. The factor \(U/4\) and the constant term in the onsite interaction are absorbed into the variational parameter. 

The onsite gates commute with each other and are applied as a single
onsite layer,
\begin{equation}
    \hat U_U(\theta_U)
    =
    \prod_i \hat U_i^{(U)}(\theta_U).
    \label{eq:hva_onsite_layer}
\end{equation}
The hopping gates are divided into commuting sets of disjoint bonds. Gates
within the same set can be applied in parallel and share one variational
parameter. This gives a compact ansatz while preserving the operator
structure of the Hubbard Hamiltonian.

We denote the commuting hopping groups by
\(\{h_g\}_{g=1}^{N_g}\), where \(N_g\) depends on the lattice geometry.
For a one-dimensional chain, nearest-neighbor hopping terms are divided
into even- and odd-bond groups, so \(N_g=2\). For a two-leg ladder, there are two horizontal hopping groups and one
vertical hopping group, so \(N_g=3\). For general two-dimensional lattices beyond the ladder geometry, the hopping terms are divided into
two horizontal and two vertical groups, so \(N_g=4\). Each hopping group defines a layer
\begin{equation}
    \hat U_g(\theta_g)
    =
    \prod_{\langle p,q\rangle\in h_g}
    \hat U_{pq}^{(t)}(\theta_g),
    \qquad
    g=1,\ldots,N_g,
    \label{eq:hva_hopping_layer}
\end{equation}
where the product is over the disjoint bonds in group $h_g$.

A single HVA layer is then written uniformly as
\begin{equation}
    \hat U_{\rm HVA}^{(\ell)}
    =
    \hat U_U(\theta_{\ell,U})
    \prod_{g=1}^{N_g}
    \hat U_g(\theta_{\ell,g}),
    \label{eq:hva_layer_general}
\end{equation}
and the full circuit with \(S\) evolution layers is
\begin{equation}
    \hat U_{\rm HVA}(\boldsymbol{\theta})
    =
    \prod_{\ell=1}^{S}
    \hat U_{\rm HVA}^{(\ell)} .
    \label{eq:hva_full}
\end{equation}
Consequently, the number of variational parameters per HVA layer is $N_{\rm par}^{\rm layer}=1+N_g$, where the additional parameter corresponds to the onsite layer. This gives
three parameters per layer for a one-dimensional chain, four parameters
per layer for a two-leg ladder, and five parameters per layer for a
general two-dimensional lattice.

Each HVA layer may equivalently be interpreted as an approximate
variational evolution under an effective Hamiltonian
\begin{equation}
    H_{\rm eff}^{(\ell)}
    =
    \theta_{\ell,U}h_U
    +
    \sum_{g=1}^{N_g}
    \theta_{\ell,g}h_g .
    \label{eq:hva_parameters}
\end{equation}
Because the parameters are optimized variationally, they are not
interpreted as fixed physical time steps. Any error associated with the
noncommuting decomposition is absorbed into the optimized angles.

The HVA preserves the fixed spin sector because both the hopping gates and
onsite gates conserve \(N_\uparrow\) and \(N_\downarrow\). Increasing the
number of layers \(S\) increases the ability of the circuit to generate
correlations beyond the non-interacting Slater determinant, but also
increases the circuit depth and the number of parameters. A 
sector-preserving structure is essential here because the VQE energies
must be computed consistently across neighboring spin sectors before
taking finite differences to obtain the spin-resolved
XC potentials.

%%%%%%%%%%%%%%%%%%%%%%%%%%%%%%%%%%%%%%%%%%%%%%%%%%%%%%%%%%%%%%%%%%%%%%%%
\begin{figure}[t]
\centering
\includegraphics[width=0.95\linewidth]{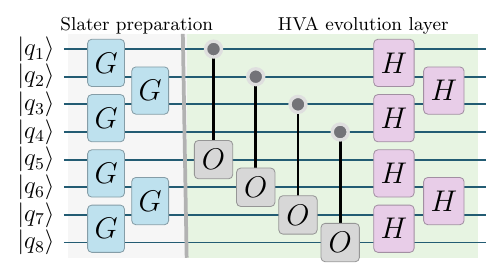}
\caption{Quantum circuit layout for a 4-site Hubbard chain, using a single HVA variational layer.
$G$: Givens rotations; 
$O$: onsite evolution gates; 
$H$: hopping evolution gates. The first four qubits encode the spin-up orbitals and the last four encode the spin-down orbitals, following the snake-ordering convention.
The onsite and hopping gates implement variational evolutions generated
by the corresponding onsite and kinetic terms of the Fermi-Hubbard
Hamiltonian. All onsite gates in a layer share the same variational parameter, the even- and odd-bond hopping layers have separate variational parameters. The layers of $G$ gates (shaded gray) are used to prepare the non-interacting ground state first. Then we use the layer of onsite and hopping gates (shaded green) to evolve the state in $U \neq 0$ regime. The resulting circuit defines the variational state used to evaluate the
energy expectation value.}
\label{fig:quantum_circ}
\end{figure}
%%%%%%%%%%%%%%%%%%%%%%%%%%%%%%%%%%%%%%%%%%%%%%%%%%%%%%%%%%%%%%%%%%%%%%%%

Figure~\ref{fig:quantum_circ} illustrates this structure for a four-site
one-dimensional Hubbard chain. The snake ordering is used,
with the first four qubits representing the spin-up orbitals and the
remaining four qubits representing the spin-down orbitals. The shaded
region on the right corresponds to a single HVA evolution layer. The
onsite gates $(O)$ couple the two spin orbitals associated with the same
lattice site, while the hopping gates $(H)$ are organized into even- and
odd-bond layers within each spin sector. Repeating this HVA block $(S)$
times yields the full variational circuit.

In the snake ordering, the two spin orbitals belonging to a given lattice
site are separated in the qubit register. Consequently, implementation on
hardware with limited qubit connectivity may require additional routing
operations, typically realized through SWAP gates, to bring interacting
qubits together~\cite{Stanisic2022,Wecker2015}. The exact routing overhead
depends on the connectivity and native gate set of the target quantum
device. Since the present work employs exact state-vector simulations, such
hardware-specific compilation costs are not considered.

\subsection{Initial-state preparation at $U=0$}
\label{sec:initial_state}

The HVA circuit described in Sec.~\ref{sec:hva} acts on an initial
reference state prepared in the desired fixed spin sector
\((N_\uparrow,N_\downarrow)\). In this work, the reference state is chosen
to be the non-interacting ground state of the Hubbard Hamiltonian at
\(U=0\). In this limit, the Hamiltonian is quadratic in the fermionic
operators and separates into independent spin sectors. Its many-body
ground state is therefore a fermionic Gaussian state, or, equivalently, a
Slater determinant obtained by occupying the lowest-energy
single-particle orbitals in each spin sector.

We prepare this non-interacting Slater determinant using the
Givens-rotation state-preparation algorithm of Jiang
et al.~\cite{Jiang2018}. Their construction provides efficient circuits
for preparing arbitrary Slater determinants and fermionic Gaussian states.
The method exploits the fact that any unitary transformation of the
single-particle orbital basis can be decomposed into a sequence of
two-mode Givens rotations, allowing the target many-body Slater
determinant to be prepared exactly from a simple reference occupation
state. Related linear-depth constructions based on
parallel nearest-neighbor Givens rotations and fermionic swap networks
were developed by Kivlichan et al.~\cite{Kivlichan2018}. This deterministic construction eliminates the need for a separate VQE optimization in the non-interacting limit.

For each spin sector, we diagonalize the non-interacting single-particle
hopping matrix. Let \(\varphi_{\alpha\sigma}(i)\) denote the
\(\alpha\)-th single-particle eigenvector for spin \(\sigma\), ordered by
increasing energy. The non-interacting many-body ground state in the sector
\((N_\uparrow,N_\downarrow)\) is
\begin{equation}
    |\Phi_0^{(N_\uparrow,N_\downarrow)}\rangle
    =
    \prod_{\alpha=1}^{N_\uparrow}
    \hat d_{\alpha\uparrow}^{\dagger}
    \prod_{\beta=1}^{N_\downarrow}
    \hat d_{\beta\downarrow}^{\dagger}
    |{\rm vac}\rangle ,
    \label{eq:noninteracting_slater}
\end{equation}
where
\begin{equation*}
    \hat d_{\alpha\sigma}^{\dagger}
    =
    \sum_i
    \varphi_{\alpha\sigma}(i)
    \hat c_{i\sigma}^{\dagger}.
\end{equation*}
The occupied single-particle orbitals define the basis transformation from
site orbitals to the occupied eigenmodes of the \(U=0\) hopping
Hamiltonian.

Any Slater determinant is completely specified by a set of orthonormal single-particle orbitals. Consequently, preparing an arbitrary ($N$-particle) Slater determinant reduces to implementing a unitary transformation in the underlying single-particle Hilbert space. Since any unitary matrix can be factorized into a sequence of two-mode Givens rotations, the corresponding many-body Slater determinant can be prepared by applying the same sequence of number-conserving fermionic rotations to a reference occupation state. Fermionic antisymmetry is preserved automatically throughout this procedure because the transformation acts linearly on the fermionic creation operators and therefore maps one Slater determinant into another~\cite{Jiang2018,Kivlichan2018}.

Following Jiang et al.~\cite{Jiang2018} and starting
from a computational-basis occupation state with the desired particle
numbers, the Givens circuit rotates the occupied site orbitals into the
occupied eigenmodes. A fermionic Givens rotation acting on modes \(p\) and
\(q\) may be written as
\begin{equation}
    \hat G_{pq}(\phi)
    =
    \exp\left[
    \phi
    \left(
    \hat c_p^\dagger \hat c_q
    -
    \hat c_q^\dagger \hat c_p
    \right)
    \right],
    \label{eq:fermionic_givens}
\end{equation}
which mixes only the two selected modes and conserves particle number.

Under a direct JWT, a two-mode fermionic rotation
between nonadjacent modes contains an intermediate parity string. However,
the preparation algorithm used here decomposes the orbital transformation
into elementary nearest-neighbor Givens rotations. For adjacent modes, the
JW string is absent and the generator reduces to a local
\(XY\)-type two-qubit operator,
\begin{equation}
    \hat c_p^\dagger \hat c_{p+1}
    -
    \hat c_{p+1}^\dagger \hat c_p
    =
    \frac{i}{2}
    \left(
    \hat X_p \hat Y_{p+1}
    -
    \hat Y_p \hat X_{p+1}
    \right),
    \label{eq:adjacent_givens_jw}
\end{equation}
up to a convention-dependent overall sign. Thus, the elementary preparation gates are local number-conserving two-mode rotations. Nonlocal orbital mixing is therefore generated through sequences of local rotations and swaps rather than through explicit long Pauli strings.
%\vspace{-0.3cm}
%\vspace{-0.2cm}
The preparation circuit can be written compactly as
\begin{equation}
    |\Phi_0^{(N_\uparrow,N_\downarrow)}\rangle
    =
    \hat U_G(\boldsymbol{\phi})
    |{\rm occ};N_\uparrow,N_\downarrow\rangle ,
    \label{eq:givens_preparation}
\end{equation}
where \(\hat U_G(\boldsymbol{\phi})\) is the product of Givens rotations
obtained from the occupied-orbital coefficient matrices, and
\(|{\rm occ};N_\uparrow,N_\downarrow\rangle\) is a reference occupation
state in the chosen spin sector. Since the \(U=0\) hopping Hamiltonian
does not couple spin-up and spin-down modes, the Givens circuits for the
two spin sectors are constructed independently and then embedded into the
chosen orbital ordering.

The left shaded region of Fig.~\ref{fig:quantum_circ} shows the
corresponding Slater-preparation block for a four-site chain. The \(G\)
gates denote the number-conserving Givens rotations used to transform the
computational-basis occupation state into the non-interacting
single-particle eigenbasis before the HVA evolution is applied.

In the standard Givens decomposition used here, an \(N_\sigma\)-particle
Slater determinant on \(L\) spatial sites is prepared using
\(N_\sigma(L-N_\sigma)\) two-mode rotations for spin \(\sigma\), before
including any additional routing or swap overhead required by the chosen
ordering and hardware connectivity. Therefore, for the spinful Hubbard
system, the number of Givens rotations used for the non-interacting
reference is
\begin{equation}
    N_G
    =
    N_\uparrow(L-N_\uparrow)
    +
    N_\downarrow(L-N_\downarrow).
    \label{eq:number_givens}
\end{equation}
The resulting state lies exactly in the desired
\((N_\uparrow,N_\downarrow)\) sector and serves as the input to the
finite-\(U\) HVA optimization described next.
\subsection{Finite-$U$ VQE optimization}
\label{sec:finite_u_vqe}

After preparing the non-interacting reference state
\(|\Phi_0^{(N_\uparrow,N_\downarrow)}\rangle\), the interacting ground
state at finite \(U\) is approximated using the HVA circuit introduced in
Sec.~\ref{sec:hva}. For a fixed spin sector
\((N_\uparrow,N_\downarrow)\), the variational state is written as
\begin{equation}
    |\Psi(\boldsymbol{\theta};U)\rangle
    =
    \hat U_{\rm HVA}(\boldsymbol{\theta})
    |\Phi_0^{(N_\uparrow,N_\downarrow)}\rangle ,
    \label{eq:vqe_state}
\end{equation}
where \(\boldsymbol{\theta}\) denotes the HVA parameters. The parameters
are optimized by minimizing the sector-resolved VQE energy
\begin{equation}
    E_{\rm VQE}^{(N_\uparrow,N_\downarrow)}(U)
    =
    \min_{\boldsymbol{\theta}}
    \langle
    \Psi(\boldsymbol{\theta};U)
    |
    \hat H(U)
    |
    \Psi(\boldsymbol{\theta};U)
    \rangle .
    \label{eq:vqe_energy}
\end{equation}
The minimization is performed independently in each fixed spin sector.
Because the HVA gates conserve \(N_\uparrow\) and \(N_\downarrow\), the
optimization remains within the target sector throughout the calculation.

We consider two finite-\(U\) optimization protocols. In the direct
protocol, the non-interacting reference state at \(U=0\) is used as the
initial state and the HVA parameters are optimized directly for the target
interaction strength \(U_{\rm target}\). This corresponds to
\begin{equation*}
    U=0
    \longrightarrow
    U_{\rm target}.
\end{equation*}
This protocol is simple and avoids intermediate calculations, but for
large \(U\) or large lattices the overlap between the non-interacting
reference state and the interacting ground state can decrease.

In the continuation protocol, the interaction strength is increased
gradually from \(U=0\) to the target value,
\begin{equation*}
    0=U_0 < U_1 < \cdots < U_M=U_{\rm target}.
\end{equation*}
At each step \(U_m\), the optimized parameters from the previous
interaction strength \(U_{m-1}\) are used as the initial parameters for
the next optimization,
$
    \boldsymbol{\theta}_{\rm init}(U_m)
    =
    \boldsymbol{\theta}^{*}(U_{m-1}) .
$
The optimized parameters at \(U_m\) are then obtained from
\begin{equation}
    \boldsymbol{\theta}^{*}(U_m)
    =
    \arg\min_{\boldsymbol{\theta}}
    \langle
    \Psi(\boldsymbol{\theta};U_m)
    |
    \hat H(U_m)
    |
    \Psi(\boldsymbol{\theta};U_m)
    \rangle .
\end{equation}
This warm-start strategy follows the interacting ground state gradually in parameter space. In the systems studied here, continuation and direct optimization exhibit similar variational-depth requirements and lead to the same overall scaling trends for the minimum number of HVA layers required to reach the target fidelity. While continuation may alter the optimal depth slightly in some sectors, it does not provide a systematic reduction in circuit depth. Moreover, because the optimization must be performed at multiple intermediate interaction strengths, the continuation approach incurs a substantially larger computational cost than direct optimization.

For both the direct and continuation protocols, we also consider a
two-stage parameter optimization. In the first stage, the parameters of
the non-interacting Slater determinant preparation circuit are kept fixed,
and only the HVA parameters are optimized. This gives the variational
state
\begin{equation}
    |\Psi_{\rm HVA}(\boldsymbol{\theta};U)\rangle
    =
    \hat U_{\rm HVA}(\boldsymbol{\theta})
    \hat U_G(\boldsymbol{\phi}_0)
    |{\rm occ};N_\uparrow,N_\downarrow\rangle ,
    \label{eq:stage1_state}
\end{equation}
where \(\hat U_G(\boldsymbol{\phi}_0)\) is the Givens-rotation circuit that
prepares the \(U=0\) Slater determinant. In the second stage, the
optimized HVA parameters from the first stage are used as the initial
point, and the full set of circuit parameters is optimized, including both
the HVA angles and the Givens-rotation angles in the Slater preparation
circuit:
\begin{equation}
    |\Psi_{\rm full}(\boldsymbol{\theta},\boldsymbol{\phi};U)\rangle
    =
    \hat U_{\rm HVA}(\boldsymbol{\theta})
    \hat U_G(\boldsymbol{\phi})
    |{\rm occ};N_\uparrow,N_\downarrow\rangle .
    \label{eq:stage2_state}
\end{equation}
In the circuit representation shown in Fig.~\ref{fig:quantum_circ}, the
state-preparation block \(\hat U_G(\boldsymbol{\phi})\) is followed by one HVA evolution layer $\hat U_{\rm HVA}(\boldsymbol{\theta})$.

The full-parameter refinement is especially useful when the
non-interacting ground state is degenerate. This is because the $U=0$ ground-state manifold may contain several Slater determinants with the same non-interacting energy but different overlaps with the interacting ground state. In nondegenerate cases, the
fixed-Slater optimization already provides a good reference and the second
stage typically gives only a modest improvement in fidelity. 

In degenerate, however, the particular Slater determinant
selected at \(U=0\) may have poor overlap with the interacting ground
state. Allowing the Givens parameters to vary lets the optimization choose
a more favorable linear combination within, or near, the non-interacting
manifold, and can significantly improve the final fidelity. We applied
this two-stage approach to assess the improvement obtained by releasing the Slater
parameters.

The quality of the optimized state is assessed by comparing the VQE energy with exact diagonalization results for system sizes where exact solutions are available. When the exact ground state $(|\Psi_{\rm exact}(U)\rangle)$ is available, we additionally compute the fidelity
\begin{equation}
\mathcal{F}
=
\left|
\langle
\Psi_{\rm exact}(U)
|
\Psi_{\rm full}(\boldsymbol{\theta}^{*},\boldsymbol{\phi}^{*};U)
\rangle
\right|^2 ,
\label{eq:fidelity}
\end{equation}
or the corresponding fixed-Slater fidelity when only the first optimization stage is employed. In the present work, the overlap fidelity serves as the primary metric for assessing the quality of state preparation, and we regard a variational state as successfully converged when $(\mathcal{F}\geq 0.99)$. The number of HVA layers is increased until this fidelity threshold is reached. For larger systems, where exact state vectors are unavailable and fidelity cannot be evaluated directly, the quality of variational states must instead be assessed using alternative diagnostics, including the convergence of the variational energy, the energy variance, the stability of physical observables with increasing circuit depth, and comparisons with approximate classical methods when available~\cite{Tilly2022,Zhang2022,Kardashin2020}. Similar validation strategies have been adopted in recent large-scale digital quantum simulations of the two-dimensional Fermi-Hubbard model, where exact classical benchmarks are available only for restricted system sizes and parameter regimes, and validation of larger systems relies on comparisons with approximate classical methods and physically motivated observables~\cite{Phasecraft2025}.

For each lattice geometry, interaction strength, and spin sector, the VQE
optimization produces the sector-resolved ground-state energy
\(E_{\rm VQE}^{(N_\uparrow,N_\downarrow)}(U)\). These energies are the
central input for the construction of the spin-resolved
XC energy and potential described in the next
subsection.

\subsection{Extraction of spin-resolved exchange-correlation potentials}
\label{sec:xc_extraction}

After the finite-\(U\) VQE optimization, we obtain an approximate
ground-state energy and optimized state in each fixed spin sector,
\[
    \left\{
    E^{(N_\uparrow,N_\downarrow)}(U),
    |\psi^{(N_\uparrow,N_\downarrow)}(U)\rangle
    \right\}.
\]
These sector-resolved energies and state vectors provide the many-body input
used to construct the lattice XC functional. For an optimized state expanded in the occupation-number basis,
$
    |\psi^{(N_\uparrow,N_\downarrow)}(U)\rangle
    =
    \sum_b c_b |b\rangle ,
$
the spin-resolved site densities are evaluated as
$
\langle n_{i\sigma}\rangle
    =
    \sum_b |c_b|^2 n_{i\sigma}(b),
$
where \(n_{i\sigma}(b)\in\{0,1\}\) is the occupation of spin
\(\sigma\) at site \(i\) in the Fock configuration \(b\). The corresponding Hartree contribution is
\begin{equation}
    E_H^{(N_\uparrow,N_\downarrow)}(U)
    =
    U\sum_i
    \langle n_{i\uparrow}\rangle
    \langle n_{i\downarrow}\rangle .
    \label{eq:hartree_energy}
\end{equation}

The XC energy in a given spin sector is then defined by
subtracting the non-interacting reference energy and the Hartree
contribution from the interacting many-body energy,
\begin{equation}
    E_{\rm XC}^{(N_\uparrow,N_\downarrow)}(U)
    =
    E^{(N_\uparrow,N_\downarrow)}(U)
    -
    E_S^{(N_\uparrow,N_\downarrow)}
    -
    E_H^{(N_\uparrow,N_\downarrow)}(U).
    \label{eq:xcenergy}
\end{equation}
Here \(E_S^{(N_\uparrow,N_\downarrow)}(\equiv T_S \text{ in Eq.~\ref{eq:energy}})\) is the non-interacting
Kohn-Sham reference energy in the same spin sector. In the present
implementation, it is approximated using the corresponding
non-interacting Hubbard reference state (\(U=0\)) described in
Sec.~\ref{sec:initial_state}. Equation~\eqref{eq:xcenergy} therefore
isolates the part of the energy that is beyond the non-interacting kinetic
energy and the mean-field Hartree interaction.

Spin-resolved XC potentials are obtained from
finite differences of \(E_{\rm XC}\) over neighboring spin sectors. For
the magnetic case, one spin population is varied while the other is kept
fixed:
\begin{equation}
\begin{aligned}
    V_{{\rm XC},\uparrow}^{(N_\uparrow,N_\downarrow)}(U)
    &=
    E_{\rm XC}^{(N_\uparrow+1,N_\downarrow)}(U)
    -
    E_{\rm XC}^{(N_\uparrow,N_\downarrow)}(U),
    \\
    V_{{\rm XC},\downarrow}^{(N_\uparrow,N_\downarrow)}(U)
    &=
    E_{\rm XC}^{(N_\uparrow,N_\downarrow+1)}(U)
    -
    E_{\rm XC}^{(N_\uparrow,N_\downarrow)}(U).
\end{aligned}
\label{eq:xcpotential_spin}
\end{equation}
Equivalently, this corresponds to changing only
\(N_\sigma\) while keeping \(N_{\bar\sigma}\) fixed.

For the nonmagnetic case, we restrict to the equal-spin line
\(N_\uparrow=N_\downarrow\). The XC potential is then evaluated along the
diagonal direction of the \((N_\uparrow,N_\downarrow)\) plane:
\begin{equation}
    V_{\rm XC}^{(N_\uparrow,N_\downarrow)}(U)
    =
    \frac{E_{\rm XC}^{(N_\uparrow+1,N_\downarrow+1)}(U)
    -
    E_{\rm XC}^{(N_\uparrow,N_\downarrow)}(U)}{2}.
    \label{eq:xcpotential_nonmagnetic}
\end{equation}
The factor of \(1/2\) converts the pair-addition energy into an effective
XC potential per added particle. This diagonal finite difference adds one spin-up and one spin-down
particle simultaneously and therefore preserves the unpolarized condition. The complete workflow is summarized in
Algorithm~\ref{alg:vqe_xc_compact}.

\begin{figure}[t]
\centering
\begin{minipage}{\columnwidth}

\rule{\linewidth}{0.4pt}
\refstepcounter{myalgorithm}
\textbf{Algorithm 2: VQE extraction of spin-resolved XC potentials}
\label{alg:vqe_xc_compact}
\rule{\linewidth}{0.4pt}

\begin{algorithmic}[1]

\For{each spin sector \((N_\uparrow,N_\downarrow)\)}

    \State Prepare the \(U=0\) Slater determinant
    \(\hat U_G(\boldsymbol{\phi}_0)|{\rm occ}\rangle\).

    \State Optimize the finite-\(U\) HVA circuit, with optional
    full-parameter refinement of both HVA and Givens angles.

    \State Store the optimized energy
    \(E^{(N_\uparrow,N_\downarrow)}(U)\) and densities
    \(\langle n_{i\sigma}\rangle\).

    \State Compute
    \(E_{\rm XC}^{(N_\uparrow,N_\downarrow)}(U)\)
    from Eq.~\eqref{eq:xcenergy}.

\EndFor

\State Obtain the magnetic and nonmagnetic XC potentials from the
finite differences in Eqs.~\eqref{eq:xcpotential_spin} and
\eqref{eq:xcpotential_nonmagnetic}.

\State Interpolate the extracted XC quantities over the spin-density plane when a smooth functional representation is needed.

\end{algorithmic}

\rule{\linewidth}{0.4pt}

\end{minipage}
\end{figure}

\section{Results}
\label{sec:results}

We benchmark the proposed quantum-classical workflow for determining spin-resolved XC quantities in the Hubbard model. The complete procedure is summarized in Algorithm~\ref{alg:vqe_xc_compact}. Unless otherwise stated, all results correspond to noiseless state-vector simulations, and the accuracy of the variational ground states and derived XC quantities is assessed through direct comparison with exact diagonalization calculations.

The calculations were performed using the Qiskit and PennyLane software frameworks. The variational parameters were optimized using the
limited-memory Broyden--Fletcher--Goldfarb--Shanno algorithm with bound constraints (L-BFGS-B)~\cite{Zhu1997} algorithm together with analytical gradients available in PennyLane. The use of analytical gradients and GPU-accelerated simulations significantly reduced the computational cost of the optimization procedure, allowing calculations for systems as large as a $3\times4$ lattice ($12$ sites). Simulations were carried out on a computer cluster using NVIDIA H100 GPUs and, for smaller calculations, on AWS GPU resources.

Throughout this work, we consider the repulsive Hubbard model with
uniform nearest-neighbor hopping $(t=1)$ and open boundary conditions (OBC). Unless
otherwise stated, all one- and two-dimensional lattices reported in
this paper employ OBC.

We first examine the quality of the variational ground-state preparation and the associated circuit-resource requirements. We then used the resulting many-body ground states to reconstruct the XC functionals and analyzed the dependence of the extracted XC potentials on particle filling, spin polarization, and lattice geometry.

\subsection{Variational ground-state preparation}

We first examine the ability of the HV ansatz to accurately prepare ground states of the Hubbard model within fixed spin sectors. The Hubbard Hamiltonian is mapped to qubits using the JWT described in Sec.~\ref{sec:fermionic}. Unless otherwise stated, the results presented in this section correspond to the snake-ordering convention. The variational circuit is initialized in the non-interacting ($U=0$) Slater determinant corresponding to the target spin sector ($N_\uparrow,N_\downarrow$), as described in Sec.~\ref{sec:initial_state}. The interaction effects are then incorporated by applying ($S$) layers of the HV ansatz.

For each spin sector and evolution depth ($S$), the VQE optimization was repeated three times using different random initialization. The fidelities, energy errors, and minimum depths reported in the main text and in Appendix \ref{app:sec_res_var_complexity} correspond to the best-performing run, selected according to the lowest variational energy. For each spin sector, the number of evolution layers was increased until a target fidelity $\mathcal{F}\ge0.99$ with respect to the exact ground state was achieved. We investigated both direct optimization from $U=0$ to the target interaction strength and continuation schemes in which the interaction was gradually increased. Although continuation can improve convergence in some cases, direct optimization from the non-interacting reference state was generally found to be more efficient for the systems considered here. After reaching the target fidelity, an additional optimization stage was performed in which all variational parameters, including the Givens-rotation parameters of the Slater preparation circuit, were re-optimized.

%%%%%%%%%%%%%%%%%%%%%%%%%%%%%%%%%%%%%%%%%%%%%%%%%%%%%%%%%%%%%%%%%%%%%%%%%%%
\begin{table}[t]
\centering
\footnotesize
\setlength{\tabcolsep}{2.1pt}
\renewcommand{\arraystretch}{1.15}

\caption{Convergence with the number of evolution layers $S$ at $U=4$ in the half-filled sector $(N_\uparrow,N_\downarrow)$. We report the relative energy error (\%) and fidelity for the direct-only (${\cal F}_{\rm d}$) and full (${\cal F}_{\rm f}$) optimization strategies.} 

\label{tab:layers_new}
\begin{tabular}{c|cccc|cccc}
\toprule
\toprule
$S$ &
$\mathrm{rel}\,\Delta E_{\rm \rm d}$ & $\mathcal{F}_{\rm d}$ &
$\mathrm{rel}\,\Delta E_{\rm f}$ & $\mathcal{F}_{\rm f}$ &
$\mathrm{rel}\,\Delta E_{\rm d}$ & $\mathcal{F}_{\rm d}$ &
$\mathrm{rel}\,\Delta E_{\rm f}$ & $\mathcal{F}_{\rm f}$ \\
\midrule

\multicolumn{9}{c}{$1\times4$ \hspace{11em} $2\times2$} \\
\midrule
2 & 2.90 & 0.9642 & 2.89 & 0.9656 & 48.67 & 0.7719 & 12.84 & 0.7719 \\
3 & 1.88 & 0.9830 & 1.55 & 0.9874 & 51.09 & 0.5109 & 0.11 & 0.9992 \\
4 & 0.00 & 1.0000 & 0.00 & 1.0000 & 51.14 & 0.5114 & 0.00 & 1.0000 \\

\midrule
\multicolumn{9}{c}{$1\times6$ \hspace{11em} $2\times3$} \\
\midrule
3 & 5.37 & 0.9289 & 3.21 & 0.93.34 & 11.91 & 0.8477 & 11.22 & 0.8480 \\
4 & 1.72 & 0.9709 & 1.37 & 0.9825 & 4.23 & 0.9043 & 2.64 & 0.9477 \\
6 & 0.58 & 0.9917 & 0.47 & 0.9928 & 2.41 & 0.9530 & 1.50 & 0.9708 \\
7 & 0.55 & 0.9908 & 0.26 & 0.9976 & 0.59 & 0.9941 & 0.24 & 0.9987 \\

\midrule
\multicolumn{9}{c}{$1\times8$ \hspace{11em} $2\times4$} \\
\midrule
4  & 2.81 & 0.9388 & 2.48 & 0.9406 & 6.62 & 0.8069 & 6.36 & 0.7991 \\
6  & 1.20 & 0.9730 & 0.61 & 0.9810 & 3.23 & 0.9129 & 2.67 & 0.9158 \\
8  & 0.38 & 0.9902 & 0.25 & 0.9950 & 1.24 & 0.9507 & 0.94 & 0.9693 \\
12 & 0.02 & 0.9997 & 0.01 & 0.9999 & 0.19 & 0.9968 & 0.15 & 0.9980 \\

\midrule
\multicolumn{9}{c}{$1\times10$ \hspace{11em} $2\times6$} \\
\midrule
5  & 3.05 & 0.9123 & 1.75 & 0.9251 & 7.34 & 0.6962 & 4.31 & 0.7813 \\
9  & 0.41 & 0.9847 & 0.33 & 0.9888 & 1.85 & 0.8934 & 1.53 & 0.9054 \\
12 & 0.12 & 0.9960 & 0.06 & 0.9990 & 0.92 & 0.9467 & 0.73 & 0.9634 \\
18 & 0.01 & 0.9998 & 0.00 & 0.9999 & 0.22 & 0.9903 & 0.17 & 0.9930 \\

\midrule
\multicolumn{9}{c}{$1\times12$ \hspace{11em} $3\times4$} \\
\midrule
6  & 2.48 & 0.9149 & 1.14 & 0.9322 & 4.35 & 0.7686 & 3.86 & 0.7884 \\
9  & 0.71 & 0.9692 & 0.42 & 0.9737 & 2.21 & 0.8634 & 1.85 & 0.8891 \\
12 & 0.19 & 0.9907 & 0.13 & 0.9951 & 1.14 & 0.9415 & 0.97 & 0.9508 \\
18 & 0.02 & 0.9995 & 0.01 & 0.9997 & 0.24 & 0.9897 & 0.22 & 0.9913 \\

\bottomrule
\end{tabular}
\end{table}
%%%%%%%%%%%%%%%%%%%%%%%%%%%%%%%%%%%%%%%%%%%%%%%%%%%%%%%%%%%%%%%%%%%%%%%%

Table~\ref{tab:layers_new} summarizes the convergence of the variational ground state as a function of the number of evolution layers (S) for several lattice geometries at $U=4$. Representative results are shown at half filling, $(N_\uparrow,N_\downarrow)=(L/2,L/2)$. The accuracy of the variational state is quantified through the relative energy error (in percentage), 
\begin{equation}
\epsilon_r =
100\times\frac{|E_{\rm VQE}-E_0|}{|E_0|},
\label{eq:error}
\end{equation}
and the fidelity defined in Eq.~(\ref{eq:fidelity}). Here, $E_0$ is exact ground state energy obtained by the exact diagonalization.  Two optimization strategies are compared. In the first strategy, only the parameters of the HV evolution layers are optimized after the direct procedure ($U=0\rightarrow4)$. In the second strategy, all variational parameters, including those associated with the Slater determinant preparation, are re-optimized. The second-stage optimization consistently improves the ground-state energy and often yields a modest improvement in fidelity, particularly when the initial solution is already close to the target fidelity. However, we find that this additional optimization does not necessarily improve the XC potentials. While the full optimization generally reduced the total energy error, it does not explicitly enforce accurate particle densities. The resulting state can therefore exhibit density errors that propagate into the XC potentials, even when the energy is improved. 
 
Consequently, the XC analysis presented below is based on the direct-only protocol unless stated otherwise.
%%%%%%%%%%%%%%%%%%%%%%%%%%%%%%%%%%%%%%%%%%%%%%%%%%%%%%%%%%%%%%%%%%%%%%%%%%%
\begin{figure}
    \centering
    \includegraphics[width=0.95\linewidth]{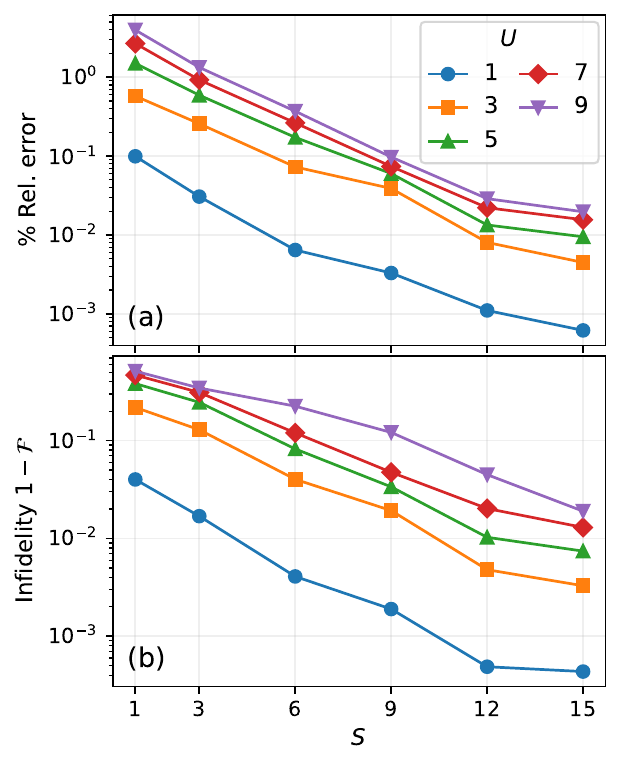}
    \caption{Convergence of the VQE solution as a function of the number of HVA evolution layers, $S$, for the ten-site Hubbard chain at half filling, $(N_\uparrow,N_\downarrow)=(5,5)$. (a) Percentage relative ground-state energy error, Eq.~(\ref{eq:error}). (b) State infidelity,
    $1-\mathcal{F}$, between the VQE state and the exact ground state. Results are shown for interaction strengths $U=1,3,5,7$ and $9$, with the curves representing the average over three optimization runs initialized from different random starting points. Solid lines are included as guides to the eye.}

    \label{fig:convergence}
\end{figure}

\begin{figure}
    \centering
    \includegraphics[width=0.95\linewidth]{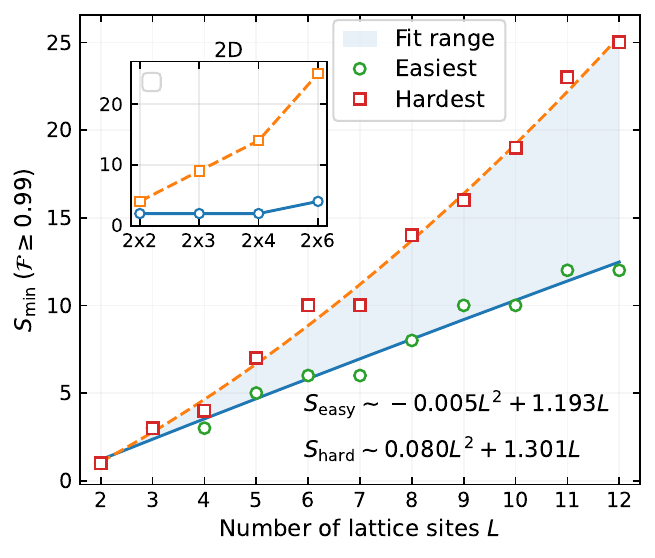}
    \caption{
Minimum number of HVA evolution layers, $S_{\min}$, required to achieve a fidelity $\mathcal{F}\ge0.99$ at $U=4$ as a function of lattice size $L$. For each system size, the easiest (green circles) and hardest (red squares) symmetry-inequivalent spin sectors are shown. Solid and dashed lines are quadratic fits to the data and serve as guides to the eye. The inset shows the corresponding results for the two-dimensional lattices $2\times2$, $2\times3$, $2\times4$, and $2\times6$. While the easiest sectors exhibit an approximately linear growth of $S_{\min}$ with system size, the hardest sectors require increasingly deeper circuits and a more pronounced quadratic dependence on $L$, reflecting the stronger correlations present in those sectors.
}
\label{fig:min_layers}
\end{figure}
%%%%%%%%%%%%%%%%%%%%%%%%%%%%%%%%%%%%%%%%%%%%%%%%%%%%%%%%%%%%%%%%%%%%%%%%%

Figure~\ref{fig:convergence} illustrates the convergence of the VQE solution with the number of evolution layers for the ten-site Hubbard chain at half filling, $(N_\uparrow,N_\downarrow)=(5,5)$. Both the relative energy error and the state infidelity decrease systematically as $S$ increases. The $\%$ relative energy error decreases systematically with increasing circuit depth, with the largest improvement occurring as the system moves from weak to intermediate interaction strength. At large $U$, the additional reduction in energy error becomes more gradual. The fidelity also improves with increasing depth, but its dependence on $U$ is less uniform, since the VQE objective is to minimize the energy rather than the wavefunction overlap. Thus, energy convergence and fidelity convergence provide complementary diagnostics of the variational state.

Reconstruction of XC potentials requires ground-state energies from multiple particle-number sectors. For a lattice containing $L$ sites, there are $(L+1)^2$ possible sectors labeled by $(N_\uparrow,N_\downarrow)$. However, for bipartite lattices, the number of independent calculations can be reduced substantially by exploiting both spin-flip and particle--hole symmetries.

The Hubbard Hamiltonian Eq.~(\ref{eq:Hubbard}) is invariant under interchange of the spin populations,
\[
(N_\uparrow,N_\downarrow)
\leftrightarrow
(N_\downarrow,N_\uparrow),
\]
which implies that sectors related by spin reversal have identical energies and XC properties. In addition, the bipartite Hubbard model possesses particle--hole symmetry, relating sectors according to
\[
(N_\uparrow,N_\downarrow)
\leftrightarrow
(L-N_\uparrow,L-N_\downarrow).
\]
Consequently, only one quarter of the full spin-sector matrix needs to be computed explicitly, while the remaining sectors can be generated using symmetry operations.

An additional reduction is possible when constructing XC potentials. The nonmagnetic XC potential is obtained along the diagonal direction $(N_\uparrow=N_\downarrow)$, and therefore only balanced-spin sectors are required. After incorporating particle--hole symmetry, only approximately half of the diagonal sectors must be calculated explicitly. For the magnetic XC potential, one spin population is kept fixed while the other is varied. In this case, particle--hole and spin-flip symmetries reduce the required calculations to approximately $(L/2)$ independent sectors for bipartite lattices. As a result, the number of VQE calculations needed for XC reconstruction is considerably smaller than the total number of particle-number sectors.

The HV layers generate the electronic correlations required to transform the non-interacting Slater determinant into the interacting ground state. The minimum number of layers required to achieve a target fidelity depends strongly on the spin sector. Figure~\ref{fig:min_layers} summarizes the minimum circuit depth ($S_{\min}$) required to reach $\mathcal{F}\ge0.99$ at $U=4$ for the easiest and hardest sectors of each lattice size.

For one-dimensional chains, the easiest sectors are generally those with low particle numbers, particularly $(N_\uparrow,N_\downarrow)=(1,1)$ for $(L\le10)$. In contrast, the most challenging sectors correspond to a single minority-spin particle moving in an almost fully polarized background, namely $(N_\uparrow,N_\downarrow)=(1,L-2)$. These sectors exhibit the largest circuit depths and show a superlinear growth of $S_{\min}$ with the size of the system. For example, the hardest sector of the $L=12$ chain requires approximately $25$ evolution layers to reach the target fidelity. The fitted curves shown in Fig.~\ref{fig:min_layers} indicate an approximately linear scaling for the easiest sectors and a noticeably faster growth in computational complexity for the hardest sectors. Complete sector-resolved results, including $S_{\min}$, fidelity, relative energy error, Hilbert-space dimension, many-body gap, Slater-state overlap, etc. are provided in Table~\ref{tab:sector_depth_gap_ent_1d} in Appendix~\ref{app:sec_res_var_complexity}.

The relative difficulty of a given spin sector also depends on the interaction strength. Although the results presented in Fig.~\ref{fig:min_layers} correspond to $U=4$, the ordering of sectors and the circuit depth required to reach a target fidelity can change as the interaction strength varies. Additional sector-resolved fidelity trends as a function of interaction strength are shown in Figs.~\ref{fig:sec_diff_1d} and ~\ref{fig:sec_diff_2d} in Appendix~\ref{app:sec_res_var_complexity}.

A different trend is observed for ladder geometries. In the $2\times3$ lattice, the sector $(2,4)$ requires $S_{\min}=9$, while for the $2\times4$ lattice, the sector $(3,5)$ requires $S_{\min}=14$. In contrast to the one-dimensional case, the most difficult sectors tend to occur closer to half filling, where the effects of electron correlation are strongest. The easiest sectors remain relatively shallow, requiring only ($S_{\min}\approx4-5$) layers even for the largest ladder considered. The most computationally demanding system studied in this work is the $3\times4$ lattice ($L=12$), which is also the largest and most complex geometry investigated. For this system, the easiest sector reaches the target fidelity with only three evolution layers, whereas the hardest sector among those examined requires approximately $30$ layers. These observations suggest that magnetization, filling, and the distance from the non-interacting reference state are more important indicators of variational complexity than the Hilbert-space dimension alone (see Table~\ref{tab:sector_depth_gap_2d} in Appendix~\ref{app:sec_res_var_complexity} ).

We also record the wall-clock cost of the VQE calculations. The runtime varies significantly across spin sectors, since some sectors converge earlier than others even at the same number of evolution layers. Most sectors considered here reach the target fidelity before $S=1.5L$. Using analytical gradients with the L-BFGS-B optimizer on a single NVIDIA H100 GPU with four vCPUs, a typical $L=10$ chain sector with $S=15$ requires approximately $8$ minutes to converge. For the $L=12$ chain, a typical sector with $S=18$ requires approximately 2 hours. The most computationally demanding calculation performed in this work was the $3\times4$ lattice $(S=30)$ in the $(5,6)$ sector, which required approximately $18$ hours on the same computational setup.

Overall, the required circuit depth increases only moderately with system size, despite the exponential growth of the Hilbert-space dimension. This behavior indicates that the HV ansatz provides an efficient representation of correlated Hubbard-model ground states for the lattice sizes investigated here and forms a suitable foundation for the extraction of XC quantities presented in the following sections.

\subsubsection*{Circuit depth and gate-count scaling}

As described in Sec.~\ref{sec:model_method}, the variational circuit consists of two components: (i) a Slater determinant preparation circuit that constructs the non-interacting ground state at $U=0$, and (ii) $S$ HV ansatz evolution layers that generate the interacting state. The non-interacting reference state is a fermionic Gaussian state prepared using Givens rotations~\cite{Jiang2018}, while the interacting part is generated by the HVA circuit~\cite{Wecker2015}. All circuits are transpiled to the gate set $\{$u, CNOT$\}$, where u is any one-qubit gate.

The complexity of the Slater determinant preparation follows directly from the fermionic Gaussian-state construction. For a lattice with $L$ sites and $N_\sigma$ fermions of spin $\sigma$, the required number of Givens rotations is given by Eq.~(\ref{eq:number_givens}). The corresponding CNOT count is
\[
N_{\mathrm{CX}}^{\mathrm{Slater}} = 
2\left[
N_\uparrow(L-N_\uparrow)
+
N_\downarrow(L-N_\downarrow)
\right].
\]
The cost is maximal near half filling, where $N_\uparrow=N_\downarrow=L/2$, yielding $N_{\mathrm{CX}}^{\mathrm{Slater}}=L^2$, while for dilute sectors such as $(1,1)$ it scales as $4(L-1)$. The two-qubit gate depth grows linearly with system size $L$. Efficient hardware implementations and resource analyses of Givens-rotation networks have been reported previously~\cite{Jiang2018,Kivlichan2018}.

The resource requirements of the HVA are controlled by the lattice size $L$ and the number of evolution layers $S$. For both one- and two-dimensional lattices, the number of two-qubit gates grows linearly with the number of evolution layers and approximately linearly with system size, yielding an overall scaling
\[
N_{2q}^{\rm HVA}  \sim O(SL).
\]
Similarly, the circuit depth scales linearly with the number of evolution layers,
\[
D_{\rm HVA} \sim O(S).
\]

The HVA consists of hopping and onsite evolution operators corresponding to the kinetic and interaction terms of the Hubbard Hamiltonian~\cite{Wecker2015}. Efficient decompositions of these operators into a constant number of two-qubit entangling gates have been demonstrated in hardware-oriented implementations of the Fermi-Hubbard model~\cite{Stanisic2022}. The exact gate counts depend on the fermion-to-qubit mapping, the basis gate set, and hardware-specific compilation strategy.

The proportionality constants depend on the lattice geometry and the fermion-to-qubit ordering convention. In particular, the snake ordering consistently yields lower gate counts and circuit depths than the zig-zag ordering. Although the zig-zag ordering renders the onsite interaction terms local, it generates longer JW strings for the hopping operators, which constitute the dominant contribution to the HVA circuit. Consequently, the increased cost of the hopping evolution outweighs the savings obtained from local onsite interactions, making the snake ordering the more resource-efficient mapping for the systems considered here.

For two-dimensional lattices, the same trend persists, with the overall circuit complexity increasing with lattice connectivity. The larger number of hopping terms and the longer JW strings required by the two-dimensional fermion-to-qubit mapping lead to higher gate counts and circuit depths than those of one-dimensional chains containing the same number of sites. Therefore, the resource requirements depend not only on the number of lattice sites and evolution layers but also on the underlying lattice geometry and ordering convention.

Combined with the variational-depth scaling discussed, these results indicate that the HVA provides a compact representation of correlated Hubbard-model ground states. Although the Hilbert-space dimension grows exponentially with system size, both the required number of evolution layers and the resulting circuit complexity increase considerably more slowly over the range of systems investigated in this work.

\subsection{Reconstruction of the XC functional}

Once accurate ground states are obtained, the corresponding
ground state energies are used to construct the
XC energy and the spin-resolved XC
potential within LDFT. The XC energy is determined by subtracting the
non-interacting kinetic energy and the Hartree contribution
from the interacting ground-state energy using
Eqs.~\eqref{eq:hartree_energy} and~\eqref{eq:xcenergy}.
The spin-resolved XC potentials are then obtained from finite
differences of the XC energy, as discussed in Sec.~\ref{sec:xc_extraction}.

%%%%%%%%%%%%%%%%%%%%%%%%%%%%%%%%%%%%%%%%%%%%%%%%%%%%%%%%%%%%%%%%%%%%%%%%%%%%%
\begin{figure*}
    \centering
    \includegraphics[width=0.9\linewidth]{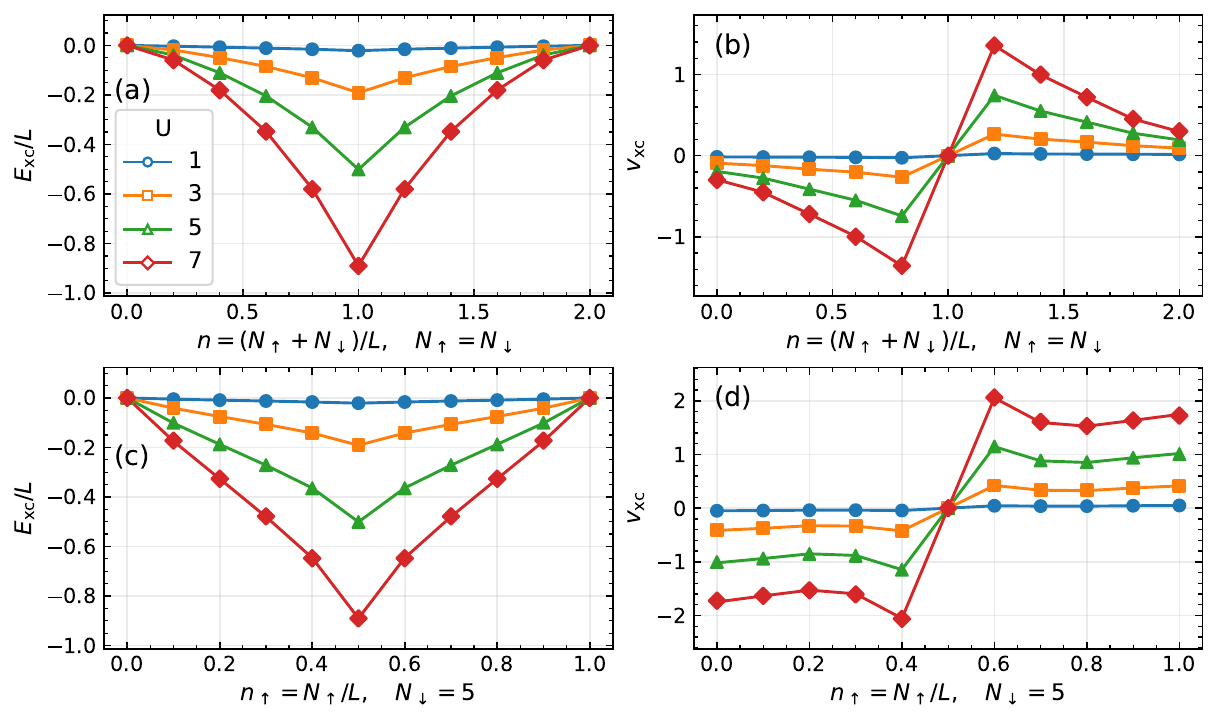}
\caption{
XC functionals obtained for the $L=10$ Hubbard chain at interaction strengths $U=1$, $3$, $5$, and $7$. Panels (a) and (b) show the XC energy per site, $\varepsilon_{\rm xc}=E_{\rm xc}/L$, and the corresponding XC potential, $v_{\rm xc}$, along the spin-balanced path $N_\uparrow=N_\downarrow$. Panels (c) and (d) show the same quantities as a function of $n_\uparrow=N_\uparrow/L$ with $N_\downarrow=5$ held fixed. Symbols denote values obtained from VQE ground states, while solid lines correspond to exact diagonalization. Increasing interaction strength enhances correlation effects, resulting in larger magnitudes of both the XC energy and the XC potential.
}
\label{fig:multiU}
\end{figure*}
%%%%%%%%%%%%%%%%%%%%%%%%%%%%%%%%%%%%%%%%%%%%%%%%%%%%%%%%%%%%%%%%%%%%%%%%%%%%

Figure~\ref{fig:multiU} illustrates the dependence of the XC functionals on the interaction strength. The XC energy is negative throughout the density range and exhibits a pronounced minimum near half filling. As the interaction strength increases, the magnitude of $\varepsilon_{\rm xc}$ grows substantially, reflecting the increasing contribution of electron correlations to the energy of the ground state. The largest changes occur in the vicinity of half filling, where the correlation effects are strongest.

The XC potential exhibits a corresponding increase in magnitude with increasing $U$. Near half filling, $v_{\rm xc}$ develops a sharp variation that becomes progressively more pronounced as the interaction strength increases. The increasingly sharp variation of $v_{\rm xc}$ near half filling is
consistent with the finite-size precursor of the derivative
discontinuity expected in LDFT and the
correlation-induced charge gap of the Hubbard model
\cite{Lima2003,Capelle2013}. The agreement between the VQE-derived and exact functionals demonstrates that accurate XC information can be reconstructed directly from variationally prepared quantum states.

\begin{figure*}
    \centering
\includegraphics[width=0.9\linewidth]{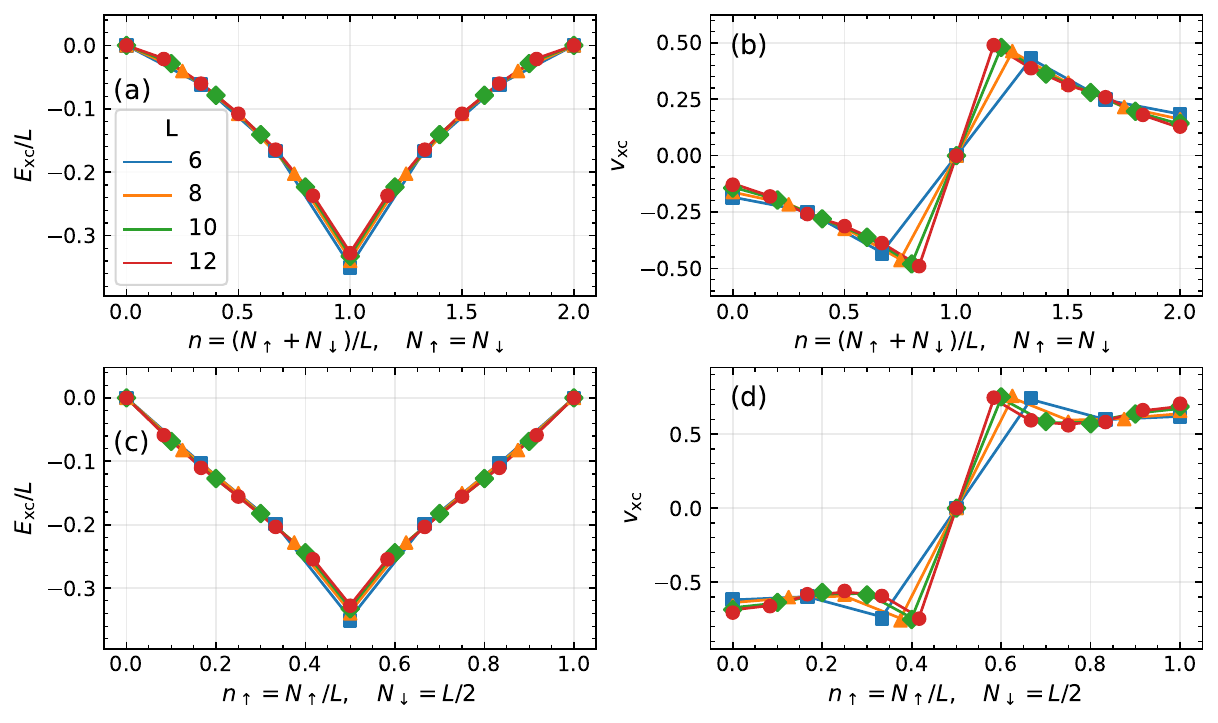}
\caption{
XC functionals for one-dimensional Hubbard chains of length $L=6$, $8$, $10$, and $12$ at $U=4$. Panels (a) and (b) show the XC energy per site and XC potential along the spin-balanced path $N_\uparrow=N_\downarrow$. Panels (c) and (d) show the corresponding spin-resolved quantities for fixed $N_\downarrow=L/2$. Symbols denote values obtained from VQE ground states and solid lines correspond to exact diagonalization. The XC functionals exhibit only a weak dependence on system size, indicating rapid convergence toward the thermodynamic-limit behavior.
}
\label{fig:multiL}
\end{figure*}

The dependence of the XC functionals on the lattice size is shown in Fig.~\ref{fig:multiL}. Remarkably, both $\varepsilon_{\rm xc}$ and $v_{\rm xc}$ display only modest finite-size effects over the range $L=6$--$12$. The XC energy curves nearly collapse onto a single universal curve when expressed as a function of density, with the largest deviations occurring near half filling. A similar trend is observed for the XC potential.

These results suggest that the XC functional converges much more rapidly with the size of the system than the underlying many-body Hilbert space, whose dimension grows exponentially with $L$. Consequently, relatively small lattice systems already capture the dominant correlation physics required to construct lattice XC functionals.

%%%%%%%%%%%%%%%%%%%%%%%%%%%%%%%%%%%%%%%%%%%%%%%%%%%%%%%%%%%%%%%%%%%%%%%%%%%%%
%%%%%%%%%%%%%%%%%%%%%%%%%%%%%%%%%%%%%%%%%%%%%%%%%%%%%%%%%%%%%%%%%%%%%%%%%%%%

%%%%%%%%%%%%%%%%%%%%%%%%%%%%%%%%%%%%%%%%%%%%%%%%%%%%%%%%%%%%%%%%%%%%%%%%%%%%
\begin{figure*}
    \centering
\includegraphics[width=1.0\linewidth]{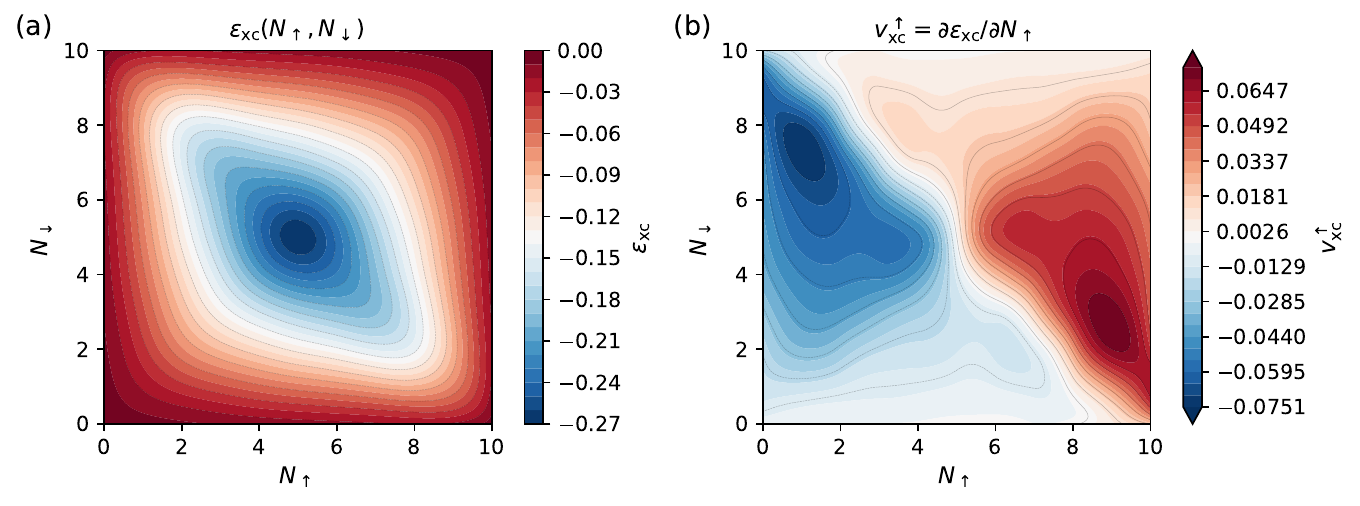}
\caption{
Two-dimensional representation of the XC functional for the $L=10$ Hubbard chain at $U=4$. (a) XC energy per site, $\varepsilon_{\rm xc}(N_\uparrow,N_\downarrow)$. (b) spin-resolved XC potential, $v_{\rm xc}^{\uparrow}=\partial \varepsilon_{\rm xc}/\partial N_\uparrow$. Values are reconstructed from VQE ground-state energies obtained in the symmetry-inequivalent spin sectors. Particle-hole and spin-flip symmetries are used to recover the full density domain, and spline interpolation is employed to obtain a smooth visualization of the functional.
}
\label{fig:contour}
\end{figure*}
%%%%%%%%%%%%%%%%%%%%%%%%%%%%%%%%%%%%%%%%%%%%%%%%%%%%%%%%%%%%%%%%%%%%%%%%%%%%

Figure~\ref{fig:contour} provides a two-dimensional view of the XC functional in the $(N_\uparrow,N_\downarrow)$ plane. The XC energy exhibits a symmetric structure arising from the combined particle-hole and spin-flip symmetries of the Hubbard Hamiltonian. The largest magnitude of $\varepsilon_{\rm xc}$ occurs near the center of the diagram, corresponding to sectors close to half filling where correlation effects are strongest. Toward the edges of the diagram, the XC energy approaches zero as the system becomes either nearly empty or nearly full.

The corresponding XC potential exhibits a more intricate structure than the XC energy. The sign of $v_{\rm xc}^{\uparrow}$ changes across the spin-balanced line and develops pronounced positive and negative regions near the half filling. This strong spin dependence demonstrates that the system cannot be adequately described by a functional of the total charge density alone. Instead, the XC potential depends sensitively on the spin composition of the ground state, reflecting the underlying many-body correlations. Consequently, the contour representation reveals that the XC functional is inherently two-dimensional in the variables $(n_\uparrow,n_\downarrow)$ and, in general, cannot be represented solely as a function of the total density $(n=n_\uparrow+n_\downarrow)$.

\begin{figure*}
    \centering
\includegraphics[width=0.9\linewidth]{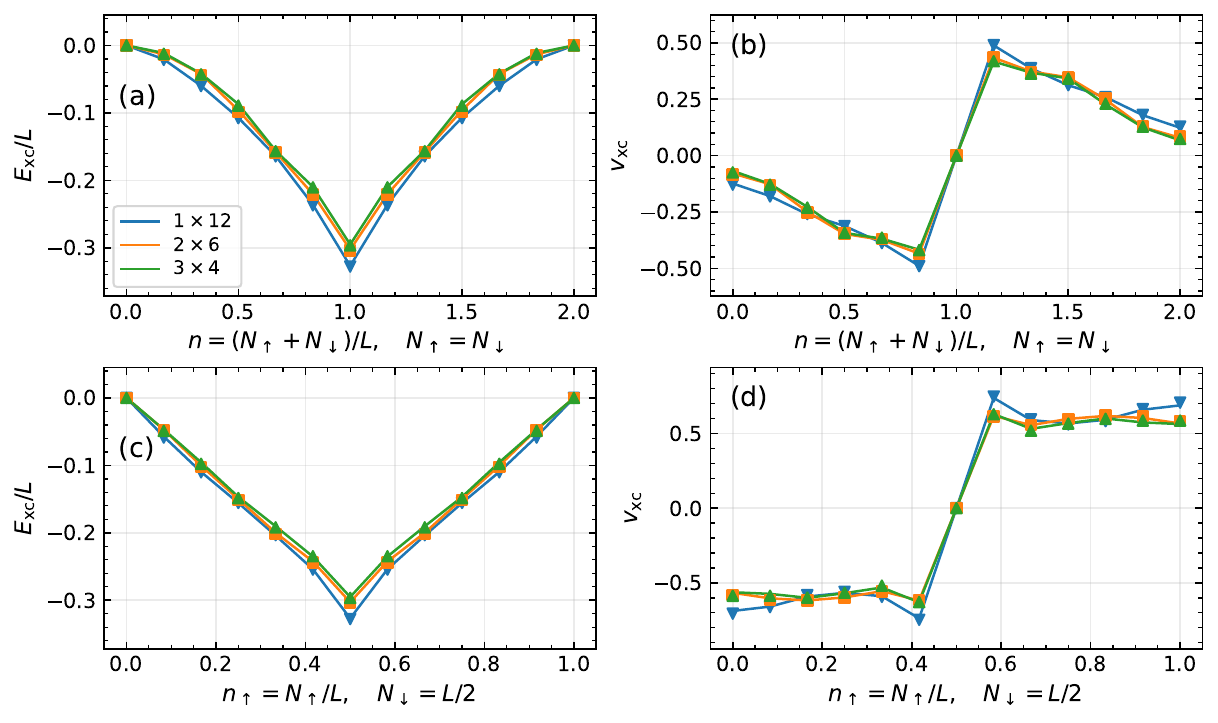}
\caption{
XC functionals for three lattice geometries containing twelve sites: a one-dimensional chain ($1\times12$), a two-leg ladder ($2\times6$), and a rectangular lattice ($3\times4$) at $U=4$. Panels (a) and (b) show the XC energy per site and XC potential along the spin-balanced path. Panels (c) and (d) show the corresponding spin-resolved quantities for fixed $N_\downarrow=L/2$. The results illustrate the influence of lattice geometry on the shape of XC functional.
}
\label{fig:exc_vxc_2D}
\end{figure*}

Figure~\ref{fig:exc_vxc_2D} compares the XC functionals obtained for three different lattice geometries that contain the same number of sites. Although the overall shape of the functional remains similar across the geometries, systematic differences are evident. The magnitude of the XC energy decreases as the geometry evolves from a one-dimensional chain to a more two-dimensional structure. The deepest minimum occurs for the $1\times12$ chain, whereas the $3\times4$ lattice exhibits the weakest XC energy.

The XC potential displays a similar geometry dependence. Although all geometries exhibit a pronounced variation near half filling, the magnitude of the potential decreases as the connectivity of the lattice increases. These results demonstrate that the XC functional depends not only on the density and interaction strength but also on the underlying lattice geometry. Consequently, lattice-specific XC functionals may be required for accurate lattice density-functional calculations in different dimensions.

Overall, these results demonstrate that variational quantum
simulations can accurately reproduce XC functionals for correlated lattice models. Because the approach relies only on ground-state preparation,
it provides a potential route for constructing improved
spin-dependent XC functionals for strongly correlated
systems using quantum simulations.

%%%%%%%%%%%%%%%%%%%%%%%%%%%%%%%%%%%%%%%%%%%%%%%%%%%%%%%%%%%%%%%%%%%%%%%%%%%%
%%%%%%%%%%%%%%%%%%%%%%%%%%%%%%%%%%%%%%%%%%%%%%%%%%%%%%%%%%%%%%%%%%%%%%%%%%%

\subsection{Spin dependence of the XC potential}

The spin dependence of the XC functional can be
analyzed more directly by examining the difference between the spin
components of the XC potential,
\[
\Delta v_{\rm XC}=v_{\rm XC}^{\uparrow}-v_{\rm XC}^{\downarrow}.
\]
This quantity measures the effective XC splitting
between spin channels and provides insight into magnetic correlation
effects in the system.

\begin{figure}[ht]
    \centering
    \includegraphics[width=0.9\linewidth]{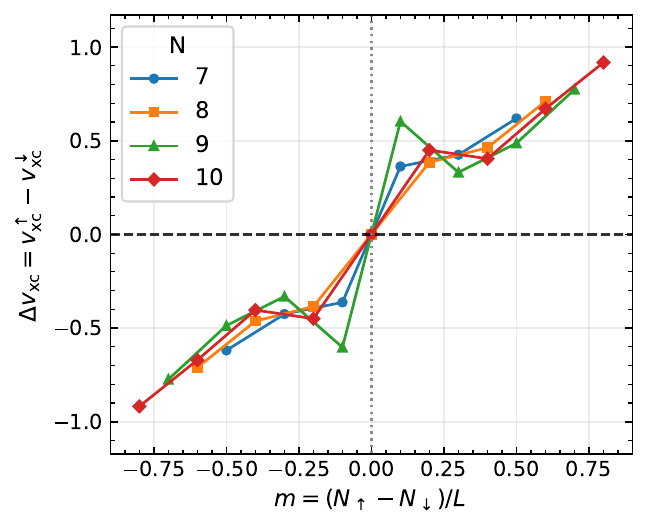}
    \caption{Spin splitting of the XC potential
    $\Delta v_{\rm XC}=v_{\rm XC}^{\uparrow}-v_{\rm XC}^{\downarrow}$
    as a function of magnetization
    $m=n_\uparrow-n_\downarrow$
    for lattice size $L=10$ at $U=4$.
    Each line corresponds to a fixed total filling
    $N=N_\uparrow+N_\downarrow$.
    The splitting vanishes in the non-magnetic sector
    and generally increases with the magnetization increase.}\label{fig:dvxc_vs_m}
\end{figure}

Figure~\ref{fig:dvxc_vs_m} shows $\Delta v_{\rm XC}$ as a function of
magnetization $m=n_\uparrow-n_\downarrow$ for a one dimensional chain of ten-sites. For the non-magnetic case ($m=0$) the two spin channels are
equivalent and the XC potentials are identical, resulting in
$\Delta v_{\rm XC}=0$. 

Although the overall magnitude of the spin splitting increases near
half filling, the curves do not show a strictly monotonic dependence on
the total filling \(N=N_\uparrow+N_\downarrow\). In particular, the
\(N=9\) curve exceeds the \(N=10\) curve at some magnetizations. This
does not imply a larger XC energy at \(N=9\); rather, it reflects the
fact that \(\Delta v_{\rm xc}\) is obtained from finite differences of
the XC energy surface and is therefore sensitive to the local curvature
of \(E_{\rm xc}(N_\uparrow,N_\downarrow)\). For the finite chains considered here, finite-size effects associated with the discrete single-particle spectrum and the discrete set of allowed spin sectors can produce such nonmonotonic behavior.

%%%%%%%%%%%%%%%%%%%%%%%%%%%%%%%%%%%%%%%%%%%%%%%%%%%%%%%%%%%%%%%%%%%%%%%%%%%
%%%%%%%%%%%%%%%%%%%%%%%%%%%%%%%%%%%%%%%%%%%%%%%%%%%%%%%%%%%%%%%%%%%%%%%%%%%

This behavior reflects the importance of spin-dependent correlation effects in the Hubbard model. As the system becomes spin polarized, spin-up and spin-down electrons experience different correlation environments, leading to an increasing separation of the two XC potentials. The overall dependence of $\Delta v_{\mathrm{XC}}$ on magnetization demonstrates that the reconstructed XC functional captures the expected evolution from non-magnetic to magnetic regimes. The approximately antisymmetric behavior about $m=0$ follows directly from the spin-flip symmetry of the Hubbard Hamiltonian. Together, these results show that the variational quantum simulations accurately reproduce not only the ground-state energies but also the spin-resolved XC effects encoded in the XC potential.

\section{Conclusion}
\label{sec:conclusion}

We have presented a quantum-algorithmic framework for constructing
spin-resolved XC energies and potentials for
the Hubbard model using variational quantum simulations. Ground states
were prepared in fixed spin sectors $(N_\uparrow,N_\downarrow)$ using a
Slater determinant reference state followed by HVA evolution layers. Although continuation in the interaction
strength was explored, we found that directly optimizing from the
non-interacting reference state at $U=0$ to the target interaction
strength provides a simple and efficient protocol for the systems
studied here.

Using noiseless state-vector simulations, we benchmarked the ansatz for
one-dimensional chains up to 12 sites and two-dimensional geometries up
to a $3\times4$ lattice. The number of HVA layers required to reach
high fidelity strongly depends on the spin sector. In one dimension,
the easiest sectors are typically dilute or weakly correlated sectors,
whereas the hardest sectors occur in highly spin-imbalanced
configurations. In contrast, for two-dimensional geometries, the most
demanding sectors tend to occur closer to half filling, where
correlation effects are strongest. These results show that
magnetization, filling, many-body gap, and the overlap with the
non-interacting Slater determinant are more informative indicators of
variational complexity than the Hilbert-space dimension alone.

From the variational ground-state energies we reconstructed
spin-resolved XC energies and finite-difference XC potentials within
LDFT. Spin-flip and particle-hole
symmetries were used to reduce the number of independent spin sectors
needed to reconstruct the full density domain. The resulting
VQE-derived XC functionals agree closely with exact diagonalization
benchmarks over the lattice sizes, interaction strengths, and spin
sectors considered. The XC energy develops its largest magnitude near
half filling, while the XC potential shows a sharp variation consistent
with the finite-size precursor of the derivative discontinuity expected
in LDFT.

We also examined the dependence of the XC functional on system size and geometry. For one-dimensional chains, the XC energy per site and the spin-resolved XC potential show only weak finite-size dependence over the range $6$ to $12$ sites. Comparing lattices with the same number of sites, including $1\times12$, $2\times6$, and $3\times4$, shows that the XC functional is geometry dependent: increasing lattice connectivity reduces the magnitude of the XC energy and modifies the structure of the XC potential. These results indicate that quantum-generated XC data can capture both spin dependence and lattice-geometry dependence in strongly correlated systems.

Finally, the circuit complexity of the HVA scales linearly with the number of evolution layers and approximately linearly with the system size, with geometry- and ordering-dependent prefactors. Since the number of layers required to reach high fidelity grows much more slowly than the many-body Hilbert-space dimension, the approach remains manageable for the lattice sizes considered here. Future work will extend this framework to larger lattice systems and improved interpolation strategies for continuous XC functionals. An important direction will be the development of analytical parameterizations of spin-resolved XC functionals from quantum-generated data, enabling their direct incorporation into LDFT calculations. In addition, the behavior of the reconstructed XC potentials near half filling may provide further insight into derivative discontinuities and correlation-driven metal-insulator transitions in strongly correlated systems. Ultimately, extending these methods to noisy quantum hardware will require measurement reduction, error mitigation, and hardware-aware circuit compilation strategies.

\section{Acknowledgments}
\label{sec-acknowledgments}

This work was supported in part by an internal seed grant from the UCF Office of Research. The authors acknowledge Amazon Web Services (AWS) for providing quantum computing simulation credits through the Amazon Braket service.

\FloatBarrier
\bibliography{references}

\appendix

\clearpage
\onecolumngrid

\section{Projection onto Fixed Particle-Number Sectors}
\label{appendix:projection}

An alternative way to restrict simulations to a specific particle
sector is to project the Hamiltonian directly onto the subspace
with a fixed number of electrons. Since the second-quantized
Hamiltonian is block diagonal in particle number,
\begin{equation}
\hat{N}_{\text{tot}}=\sum_{j=1}^{K} \hat{n}_j ,
\end{equation}
one can isolate the desired $N$-electron sector using the projector
\begin{equation}
\hat{P}_N=\prod_{j\neq N}\frac{\hat{N}_{\text{tot}}-j\hat{I}}{N-j},
\end{equation}
where $\hat{I}$ is the identity operator. Applying the projection operator yields
\begin{equation}
\hat{H}_N = \hat{P}_N\, \hat{H}\, \hat{P}_N ,
\end{equation}
which acts only on the Hilbert space corresponding to the
$N$-particle sector. This reduces the Hilbert space dimension from
$2^{K}$ to $\dim\mathcal{H}_N=\binom{K}{N}$. Further reduction can be obtained by fixing the spin sector
$(N_\uparrow,N_\downarrow)$, leading to
\begin{equation}
\dim\mathcal{H}_{N_\uparrow,N_\downarrow}
= \binom{K/2}{N_\uparrow}\binom{K/2}{N_\downarrow}.
\end{equation}

The projector can be applied directly in operator space using the
fermionic anticommutation relations without constructing the full
matrix representation of the Hamiltonian~\cite{Moll2016}.
Subsequent qubit reductions can be performed iteratively using
shift operators,
\begin{equation}
\hat{I} \otimes \hat{H}^{(K-1)} = \hat{S}_{+}^{K}\, \hat{H}^{(K)} \hat{S}_{-}^{K} + \hat{S}_{-}^{K}\, \hat{H}^{(K)}\, \hat{S}_{+}^{K},
\end{equation}
with
\begin{equation}
\hat{S}_{\pm}^{(K)} = \frac{1}{2}\left( \hat{X}_{K-1} \mp i \hat{Y}_{K-1} + \hat{X}_{K-1}\, \hat{X}_{K} \pm i \hat{Y}_{K-1}\, \hat{Y}_{K} \right).
\end{equation}

As an illustrative example, the two-site Hubbard model
($K=4$ spin orbitals) can be reduced to a two-qubit Hamiltonian
in the sector $N_\uparrow=N_\downarrow=1$,
\begin{equation}
\hat{H}^{2} = -t \left( \hat{X} \otimes \hat{I} + \hat{I} \otimes \hat{X}\right) + \frac{U}{2} \left( \hat{I} + \hat{Z} \otimes \hat{Z} \right).
\end{equation}
While this projection approach reduces the number of qubits,
the resulting Hamiltonian generally contains many more Pauli
terms, which increases circuit depth and measurement cost.
For this reason we do not employ the projection method in the
main simulations presented in this work.

\section{Sector-resolved variational complexity}
\label{app:sec_res_var_complexity}

To quantify the complexity of variational ground-state preparation across the Hubbard-model Hilbert space, we analyzed the minimum number of HVA evolution layers required to reach a target fidelity of $\mathcal{F}\ge0.99$ in each spin sector. The appendix provides sector-resolved results for all symmetry-inequivalent sectors considered in this work, including the minimum circuit depth $S_{\min}$, achieved fidelity, relative energy error, Hilbert-space dimension, many-body gap, entanglement entropy (for one-dimensional systems), magnetization, and the overlap between the non-interacting Slater determinant and the interacting ground state. These data reveal substantial variations in variational complexity across sectors and provide insight into the physical factors governing the efficiency of the HVA representation.

For the one-dimensional systems, the entanglement entropy reported in Table~\ref{tab:sector_depth_gap_ent_1d} is computed from the von Neumann entropy of the reduced density matrix obtained by partitioning the lattice into two equal halves. This bipartition provides a natural measure of the quantum correlations present in the ground state and is used consistently for all one-dimensional chains considered in this work. For the two-dimensional lattices, no entanglement entropy values are reported in Table~\ref{tab:sector_depth_gap_2d}. Unlike the one-dimensional case, there is no unique way to partition a two-dimensional lattice into subsystems, and different bipartitions (e.g., vertical, horizontal, diagonal, or irregular cuts) generally yield different entanglement entropies. To avoid introducing an arbitrary choice of partition, we restrict the entanglement analysis to the one-dimensional systems and omit this quantity for the two-dimensional geometries.

\begingroup
\small
\setlength{\tabcolsep}{4.5pt}
\renewcommand{\arraystretch}{1.10}
\begin{longtable}{ccccrrrrrr}
\caption{
Sector-resolved variational and exact-state diagnostics for symmetry-inequivalent Hubbard-model sectors at $U=4$. For each spin sector $(N_\uparrow,N_\downarrow)$, the table lists the magnetization $m=N_\downarrow-N_\uparrow$, the minimum number of HVA evolution layers $S_{\min}$ required to reach fidelity $\mathcal{F_{\rm d}}>0.99$, the achieved fidelity $\mathcal{F_{\rm d}}$, the relative energy error $\epsilon_r=|E_{\rm VQE}-E_0|/|E_0|$, the Hilbert-space dimension $D$, the many-body gap $\Delta=E_1-E_0$, the entanglement entropy $S_{\rm ent}$, and the overlap between the non-interacting and interacting ground states, $F_{\rm Slater}=|\langle \Psi(U=0)|\Psi(U)\rangle|^2$. Among the three random initializations used for each sector, only the best result is reported.
}
\label{tab:sector_depth_gap_ent_1d}
\\
\toprule
{$N_\uparrow$} & {$N_\downarrow$} & $m$ & {$S_{\min}$} & {$\mathcal{F}_{\rm d}$} & {$\epsilon_r$}  & {$D$} & {$\Delta$} & {$S_{\rm ent}$} & {$\mathcal{F}_{\rm Slater}$}\\
\midrule
\endfirsthead
\multicolumn{10}{c}{\tablename\ \thetable\ -- continued from previous page}\\
\toprule
{$N_\uparrow$} & {$N_\downarrow$} & $m$ & {$S_{\min}$} & {$\mathcal{F}_{\rm d}$} & {$\epsilon_r$} & {$D$} & {$\Delta$} & {$S_{\rm ent}$} & {$\mathcal{F}_{\rm Slater}$}\\
\midrule
\endhead
\midrule
\multicolumn{10}{r}{Continued on next page}\\
\endfoot
\bottomrule
\endlastfoot
\multicolumn{10}{c}{\textbf{$1\times 4$}}\\
\midrule
1 & 1 & 0 & 3 & 1.000 & 6.9e-7 & 16 & 0.389 & 1.80 & 0.873\\
1 & 2 & 1 & 3 & 0.992 & 9.5e-3 & 24 & 0.625 & 1.60 & 0.853\\
1 & 3 & 2 & 3 & 0.999 & 2.6e-3 & 16 & 0.553 & 1.59 & 0.762\\
2 & 2 & 0 & 4 & 1.000 & 2.5e-6 & 36 & 0.540 & 0.86 & 0.716\\
\addlinespace[2pt]
\midrule
\multicolumn{10}{c}{\textbf{$1\times 5$}}\\
\midrule
1 & 1 & 0 & 5 & 0.998 & 1.3e-3 & 25 & 0.263 & 1.76 & 0.866\\
1 & 2 & 1 &6 & 0.992 & 5.9e-3 & 50 & 0.466 & 1.55 & 0.862\\
1 & 3 & 2 &6 & 0.991 & 6.2e-3 & 50 & 0.473 & 1.76 & 0.821\\
1 & 4 & 3 &6 & 0.996 & 2.7e-3 & 25 & 0.392 & 1.37 & 0.709\\
2 & 2 & 0 &7 & 0.991 & 6.7e-3 & 100 & 0.452 & 1.62 & 0.769\\
2 & 3 & 1 &6 & 0.992 & 4.5e-3 & 100 & 0.568 & 1.08 & 0.670\\
\addlinespace[2pt]
\midrule
\multicolumn{10}{c}{\textbf{$1\times 6$}}\\
\midrule
1 & 1 & 0 & 6 & 0.999 & 1.2e-3 & 36 & 0.184 & 1.80 & 0.857\\
1 & 2 & 1 & 6 & 0.991 & 4.3e-3 & 90 & 0.348 & 1.63 & 0.860\\
1 & 3 & 2 & 7 & 0.996 & 2.4e-3 & 120 & 0.376 & 1.89 & 0.840\\
1 & 4 & 3 & 10 & 0.996 & 2.7e-3 & 90 & 0.354 & 1.82 & 0.785\\
1 & 5 & 4 & 6 & 0.995 & 2.3e-3 & 36 & 0.286 & 1.50 & 0.656\\
2 & 2 & 0 & 7 & 0.991 & 3.4e-3 & 225 & 0.359 & 1.41 & 0.782\\
2 & 3 & 1 & 7 & 0.996 & 3.5e-3 & 300 & 0.495 & 1.68 & 0.738\\
2 & 4 & 2 & 7 & 0.990 & 1.4e-2 & 225 & 0.456 & 0.82 & 0.617\\
3 & 3 & 0 & 6 & 0.992 & 5.8e-3 & 400 & 0.401 & 1.63 & 0.594\\
\addlinespace[2pt]
\midrule
\multicolumn{10}{c}{\textbf{$1\times 7$}}\\
\midrule
1 & 1 & 0 & 6 & 0.991 & 4.6e-3 & 49 & 0.133 & 1.78 & 0.847\\
1 & 2 & 1 & 8 & 0.991 & 3.5e-3 & 147 & 0.264 & 1.61 & 0.855\\
1 & 3 & 2 & 9 & 0.992 & 2.7e-3 & 245 & 0.297 & 1.85 & 0.845\\
1 & 4 & 3 & 10 & 0.993 & 2.1e-3 & 245 & 0.295 & 1.79 & 0.815\\
1 & 5 & 4 & 10 & 0.992 & 3.5e-3 & 147 & 0.269 & 1.86 & 0.748\\
1 & 6 & 5 & 7 & 0.991 & 1.1e-2 & 49 & 0.216 & 1.38 & 0.607\\
2 & 2 & 0 & 9 & 0.990 & 2.6e-3 & 441 & 0.282 & 1.61 & 0.784\\
2 & 3 & 1 & 8 & 0.990 & 3.8e-3 & 735 & 0.412 & 1.58 & 0.760\\
2 & 4 & 2 & 10 & 0.992 & 2.7e-3 & 735 & 0.412 & 1.67 & 0.699\\
2 & 5 & 3 & 8 & 0.992 & 7.6e-3 & 441 & 0.361 & 1.01 & 0.562\\
3 & 3 & 0 & 10 & 0.991 & 3.1e-3 & 1225 & 0.358 & 1.91 & 0.669\\
3 & 4 & 1 & 8 & 0.993 & 5.2e-3 & 1225 & 0.439 & 1.30 & 0.551\\
\addlinespace[2pt]
\midrule
\multicolumn{10}{c}{\textbf{$1\times 8$}}\\
\midrule
1 & 1 & 0 & 8 & 0.999 & 7.7e-4 & 64 & 0.098 & 1.79 & 0.838\\
1 & 2 & 1 & 9 & 0.990 & 2.9e-3 & 224 & 0.203 & 1.65 & 0.848\\
1 & 3 & 2 & 9 & 0.992 & 2.3e-3 & 448 & 0.236 & 1.93 & 0.844\\
1 & 4 & 3 & 9 & 0.992 & 2.6e-3 & 560 & 0.242 & 1.85 & 0.826\\
1 & 5 & 4 & 12 & 0.991 & 2.7e-3 & 448 & 0.233 & 1.93 & 0.787\\
1 & 6 & 5 & 14 & 0.992 & 2.3e-3 & 224 & 0.209 & 1.92 & 0.710\\
1 & 7 & 6 & 9 & 0.992 & 4.9e-3 & 64 & 0.168 & 1.43 & 0.562\\
2 & 2 & 0 & 10 & 0.992 & 1.6e-3 & 784 & 0.224 & 1.51 & 0.780\\
2 & 3 & 1 & 9 & 0.991 & 2.4e-3 & 1568 & 0.339 & 1.71 & 0.767\\
2 & 4 & 2 & 9 & 0.991 & 2.6e-3 & 1960 & 0.354 & 1.43 & 0.730\\
2 & 5 & 3 & 13 & 0.996 & 1.5e-3 & 1568 & 0.336 & 1.74 & 0.657\\
2 & 6 & 4 & 9 & 0.991 & 8.4e-3 & 784 & 0.288 & 0.85 & 0.509\\
3 & 3 & 0 & 10 & 0.991 & 2.4e-3 & 3136 & 0.307 & 1.98 & 0.695\\
3 & 4 & 1 & 9 & 0.992 & 2.7e-3 & 3920 & 0.400 & 1.84 & 0.636\\
3 & 5 & 2 & 9 & 0.993 & 6.1e-3 & 3136 & 0.376 & 1.64 & 0.502\\
4 & 4 & 0 & 8 & 0.990 & 3.9e-3 & 4900 & 0.319 & 1.13 & 0.489\\
\addlinespace[2pt]
\midrule
\multicolumn{10}{c}{\textbf{$1\times 9$}}\\
\midrule
1 & 1 & 0 & 10 & 0.993 & 2.3e-3 & 81 & 0.075 & 1.77 & 0.829\\
1 & 2 & 1 & 12 & 0.995 & 1.5e-3 & 324 & 0.160 & 1.63 & 0.841\\
1 & 3 & 2 & 10 & 0.991 & 2.1e-3 & 756 & 0.190 & 1.89 & 0.840\\
1 & 4 & 3 & 11 & 0.991 & 2.0e-3 & 1134 & 0.199 & 1.86 & 0.829\\
1 & 5 & 4 & 12 & 0.991 & 2.1e-3 & 1134 & 0.198 & 1.92 & 0.805\\
1 & 6 & 5 & 13 & 0.990 & 2.1e-3 & 756 & 0.187 & 1.87 & 0.759\\
1 & 7 & 6 & 16 & 0.990 & 2.3e-3 & 324 & 0.166 & 1.93 & 0.672\\
1 & 8 & 7 & 12 & 0.991 & 8.7e-3 & 81 & 0.134 & 1.35 & 0.522\\
2 & 2 & 0 & 12 & 0.992 & 1.5e-3 & 1296 & 0.179 & 1.61 & 0.774\\
2 & 3 & 1 & 10 & 0.992 & 2.0e-3 & 3024 & 0.279 & 1.67 & 0.768\\
2 & 4 & 2 & 11 & 0.991 & 2.1e-3 & 4536 & 0.299 & 1.67 & 0.743\\
2 & 5 & 3 & 13 & 0.992 & 1.5e-3 & 4536 & 0.296 & 1.60 & 0.697\\
2 & 6 & 4 & 13 & 0.991 & 2.3e-3 & 3024 & 0.274 & 1.73 & 0.614\\
2 & 7 & 5 & 11 & 0.991 & 8.6e-3 & 1296 & 0.233 & 0.95 & 0.459\\
3 & 3 & 0 & 11 & 0.991 & 2.1e-3 & 7056 & 0.260 & 1.97 & 0.706\\
3 & 4 & 1 & 12 & 0.991 & 2.1e-3 & 10584 & 0.352 & 1.77 & 0.669\\
3 & 5 & 2 & 13 & 0.992 & 1.7e-3 & 10584 & 0.349 & 1.94 & 0.595\\
3 & 6 & 3 & 10 & 0.990 & 6.8e-3 & 7056 & 0.317 & 1.43 & 0.453\\
4 & 4 & 0 & 14 & 0.991 & 1.6e-3 & 15876 & 0.295 & 1.91 & 0.576\\
4 & 5 & 1 & 10 & 0.991 & 6.4e-3 & 15876 & 0.357 & 1.28 & 0.450\\
\addlinespace[2pt]
\midrule
\multicolumn{10}{c}{\textbf{$1\times 10$}}\\
\midrule
1 & 1 & 0 & 12 & 0.998 & 9.5e-4 & 100 & 0.058 & 1.77 & 0.820\\
1 & 2 & 1 & 12 & 0.992 & 2.1e-3 & 450 & 0.127 & 1.66 & 0.833\\
1 & 3 & 2 & 12 & 0.992 & 1.5e-3 & 1200 & 0.154 & 1.94 & 0.834\\
1 & 4 & 3 & 13 & 0.992 & 1.3e-3 & 2100 & 0.165 & 1.88 & 0.828\\
1 & 5 & 4 & 10 & 0.990 & 2.0e-3 & 2520 & 0.167 & 2.00 & 0.813\\
1 & 6 & 5 & 14 & 0.993 & 1.4e-3 & 2100 & 0.163 & 1.94 & 0.783\\
1 & 7 & 6 & 15 & 0.991 & 1.8e-3 & 1200 & 0.152 & 1.93 & 0.730\\
1 & 8 & 7 & 19 & 0.991 & 2.0e-3 & 450 & 0.134 & 1.98 & 0.636\\
1 & 9 & 8 & 12 & 0.991 & 1.1e-2 & 100 & 0.109 & 1.37 & 0.486\\
2 & 2 & 0 & 13 & 0.992 & 1.3e-3 & 2025 & 0.145 & 1.55 & 0.766\\
2 & 3 & 1 & 10 & 0.990 & 1.8e-3 & 5400 & 0.231 & 1.73 & 0.764\\
2 & 4 & 2 & 13 & 0.992 & 1.2e-3 & 9450 & 0.253 & 1.55 & 0.748\\
2 & 5 & 3 & 11 & 0.991 & 2.1e-3 & 11340 & 0.256 & 1.77 & 0.716\\
2 & 6 & 4 & 14 & 0.991 & 1.8e-3 & 9450 & 0.248 & 1.49 & 0.663\\
2 & 7 & 5 & 15 & 0.990 & 1.8e-3 & 5400 & 0.226 & 1.77 & 0.571\\
2 & 8 & 6 & 12 & 0.991 & 5.7e-3 & 2025 & 0.192 & 0.85 & 0.414\\
3 & 3 & 0 & 13 & 0.992 & 1.1e-3 & 14400 & 0.219 & 2.03 & 0.708\\
3 & 4 & 1 & 12 & 0.990 & 1.6e-3 & 25200 & 0.305 & 1.86 & 0.684\\
3 & 5 & 2 & 12 & 0.994 & 1.4e-3 & 30240 & 0.313 & 2.01 & 0.636\\
3 & 6 & 3 & 13 & 0.991 & 2.7e-3 & 25200 & 0.299 & 1.98 & 0.554\\
3 & 7 & 4 & 13 & 0.995 & 1.9e-3 & 14400 & 0.267 & 1.61 & 0.406\\
4 & 4 & 0 & 14 & 0.991 & 1.6e-3 & 44100 & 0.263 & 1.70 & 0.613\\
4 & 5 & 1 & 12 & 0.994 & 1.5e-3 & 52920 & 0.333 & 1.90 & 0.543\\
4 & 6 & 2 & 12 & 0.995 & 2.9e-3 & 44100 & 0.318 & 1.06 & 0.408\\
5 & 5 & 0 & 11 & 0.991 & 2.2e-3 & 63504 & 0.266 & 1.67 & 0.399\\
\addlinespace[2pt]
\midrule
\multicolumn{10}{c}{\textbf{$1\times 11$}}\\
\midrule
1 & 1 & 0 & 13 & 0.996 & 1.8e-3 & 121    & 0.046 & 1.76 & 0.813 \\
1 & 3 & 2 & 13 & 0.990 & 1.6e-3 & 1815   & 0.127 & 1.92 & 0.828 \\
1 & 9 & 8 & 23 & 0.991 & 1.6e-3 & 605    & 0.110 & 1.98 & 0.601 \\
2 & 3 & 1 & 13 & 0.991 & 1.5e-3 & 9075   & 0.193 & 1.71 & 0.759 \\
3 & 4 & 1 & 13 & 0.992 & 1.3e-3 & 54450  & 0.264 & 1.85 & 0.690 \\
3 & 8 & 5 & 13 & 0.993 & 3.3e-3 & 27225  & 0.226 & 1.46 & 0.363 \\
4 & 7 & 3 & 12 & 0.992 & 3.7e-3 & 108900 & 0.278 & 1.25 & 0.366 \\
5 & 6 & 1 & 12 & 0.992 & 3.5e-3 & 213444 & 0.300 & 1.39 & 0.366 \\
\addlinespace[2pt]
\midrule
\multicolumn{10}{c}{\textbf{$1\times 12$}}\\
\midrule
1 & 1 & 0 & 18 & 0.995 & 6.9e-3 & 144 & 0.037 & 1.76 & 0.805\\
2 & 2 & 0  & 16 & 0.992 & 6.8e-3 & 4356 & 0.099 & 1.57 & 0.750\\
3 & 3 & 0 & 15 & 0.992 & 6.9e-3 & 48400 & 0.158 & 2.06 & 0.703\\
4 & 4 & 0 & 14 & 0.990 & 1.0e-2 & 245025 & 0.204 & 1.82 & 0.638\\
5 & 5 & 0 & 17 & 0.990 & 9.3e-3 & 627264 & 0.229 & 2.09 & 0.535\\
6 & 6 & 0 & 12 & 0.991 & 1.2e-2 & 853776 & 0.227 & 1.26 & 0.325\\
1 & 6 & 5 & 13 & 0.990 & 1.1e-2 & 11088 & 0.121 & 1.97 & 0.801\\
2 & 6 & 4 & 13 & 0.991 & 1.2e-2 & 60984 & 0.192 & 1.63 & 0.702\\
3 & 6 & 3 & 13 & 0.991 & 1.2e-2 & 203280 & 0.244 & 1.99 & 0.627\\
4 & 6 & 2 & 13 & 0.990 & 1.2e-2 & 457380 & 0.276 & 1.68 & 0.552\\
5 & 6 & 1 & 14 & 0.992 & 9.9e-3 & 731808 & 0.285 & 2.00 & 0.461\\
2 & 9 & 7 & 22 & 0.990  & 1.2e-3 & 14520 & 0.1585 & 1.78 & 0.490 \\
1 & 10 & 9 & 25 & 0.991 & 1.5e-3 & 792 & 0.092 & 2.02 & 0.569 \\
\end{longtable}
\endgroup

\begingroup
\small
\setlength{\tabcolsep}{4.5pt}
\renewcommand{\arraystretch}{1.10}
\begin{longtable}{ccccrrrrr}
\caption{
Sector-resolved variational and exact-state diagnostics for symmetry-inequivalent sectors of the two-dimensional Hubbard model at $U=4$. Results are shown for the $2\times3$, $2\times4$, $2\times6$ and $3\times4$ lattices. For each spin sector $(N_\uparrow,N_\downarrow)$, the table lists the magnetization $m=N_\downarrow-N_\uparrow$, the minimum number of HVA evolution layers $S_{\min}$ required to reach fidelity $\mathcal{F_{\rm d}}>0.99$, the achieved fidelity $\mathcal{F_{\rm d}}$, the relative energy error $\epsilon_r=|E_{\rm VQE}-E_0|/|E_0|$, the Hilbert-space dimension $D$, the many-body gap $\Delta=E_1-E_0$, and the overlap between the non-interacting Slater determinant and the interacting ground state, $\mathcal{F}_{\rm Slater}=|\langle\Psi(U=0)|\Psi(U)\rangle|^2$. The reported values correspond to the best result obtained from three independent random initializations. Compared with the one-dimensional systems, the sectors requiring the largest circuit depths are generally found near half filling, reflecting the stronger correlation effects present in two-dimensional geometries.
}

\label{tab:sector_depth_gap_2d}
\\
\toprule
{$N_\uparrow$} & {$N_\downarrow$} & {$m$} & {$S_{\min}$} & {$\mathcal{F}_{\rm d}$} & {$\epsilon_r$} & {$D$} & {$\Delta$} & {$\mathcal{F}_{\rm Slater}$}\\
\midrule
\endfirsthead

\multicolumn{9}{c}{\tablename\ \thetable\ -- continued from previous page}\\
\toprule
{$N_\uparrow$} & {$N_\downarrow$} & {$m$} & {$S_{\min}$} & {$\mathcal{F}_{\rm d}$} & {$\epsilon_r$} & {$D$} & {$\Delta$} & {$\mathcal{F}_{\rm Slater}$}\\
\midrule
\endhead

\midrule
\multicolumn{9}{r}{Continued on next page}\\
\endfoot

\bottomrule
\endlastfoot

\multicolumn{9}{c}{$2\times3$} \\
\midrule
1 & 1 & 0 & 2 & 0.990 & 6.3e-3 &   36 & 0.979 & 0.953 \\
1 & 2 & 1 & 2 & 0.991 & 6.1e-3 &   90 & 0.707 & 0.926 \\
1 & 3 & 2 & 4 & 0.995 & 4.0e-3 &  120 & 0.772 & 0.877 \\
1 & 4 & 3 & 5 & 0.994 & 5.8e-3 &   90 & 0.302 & 0.825 \\
1 & 5 & 4 & 4 & 0.999 & 7.9e-4 &   36 & 0.914 & 0.814 \\
2 & 2 & 0 & 5 & 0.992 & 3.3e-3 &  225 & 0.245 & 0.808 \\
2 & 3 & 1 & 5 & 0.993 & 4.0e-3 &  300 & 0.296 & 0.769 \\
2 & 4 & 2 & 9 & 0.995 & 5.0e-3 &  225 & 0.474 & 0.579 \\
3 & 3 & 0 & 7 & 0.994 & 5.9e-3 &  400 & 0.397 & 0.556 \\

\midrule
\multicolumn{9}{c}{$2\times4$} \\
\midrule
1 & 1 & 0 & 4  & 0.999 & 1.3e-3 &   64 & 0.661 & 0.956 \\
1 & 2 & 1 & 2  & 0.992 & 4.0e-3 &  224 & 0.875 & 0.943 \\
1 & 3 & 2 & 5  & 0.993 & 2.9e-3 &  448 & 0.095 & 0.889 \\
1 & 4 & 3 & 6  & 0.994 & 2.6e-3 &  560 & 0.511 & 0.862 \\
1 & 5 & 4 & 5  & 0.992 & 3.8e-3 &  448 & 0.384 & 0.844 \\
1 & 6 & 5 & 5  & 0.998 & 2.4e-3 &  224 & 0.641 & 0.834 \\
1 & 7 & 6 & 4  & 0.992 & 1.1e-2 &   64 & 0.649 & 0.794 \\
2 & 2 & 0 & 3  & 0.992 & 4.0e-3 &  784 & 0.800 & 0.880 \\
2 & 3 & 1 & 5  & 0.997 & 1.7e-3 & 1568 & 0.170 & 0.823 \\
2 & 4 & 2 & 6  & 0.991 & 3.4e-3 & 1960 & 0.495 & 0.782 \\
2 & 5 & 3 & 5  & 0.992 & 4.9e-3 & 1568 & 0.121 & 0.720 \\
2 & 6 & 4 & 7  & 0.991 & 3.9e-3 &  784 & 0.617 & 0.630 \\
3 & 3 & 0 & 13 & 0.991 & 1.4e-3 & 3136 & 0.119 & 0.612 \\
3 & 4 & 1 & 10 & 0.992 & 2.8e-3 & 3920 & 0.109 & 0.639 \\
3 & 5 & 2 & 14 & 0.992 & 3.2e-3 & 3136 & 0.365 & 0.384 \\
4 & 4 & 0 & 12 & 0.997 & 1.9e-3 & 4900 & 0.324 & 0.444 \\
\addlinespace[2pt]
\midrule
\multicolumn{9}{c}{\textbf{$2\times 6$}}\\
\midrule
1 & 1 & 0 & 6  & 0.991 & 2.3e-3 &    144 & 0.330 & 0.950\\
2 & 2 & 0 & 4  & 0.991 & 2.5e-3 &   4356 & 0.563 & 0.914\\
3 & 3 & 0 & 5  & 0.991 & 2.4e-3 &  48400 & 0.544 & 0.824\\
4 & 4 & 0 & 18 & 0.992 & 5.7e-4 & 245025 & 0.128 & 0.628\\
5 & 5 & 0 & 25 & 0.995 & 8.5e-4 & 627264 & 0.146 & 0.469\\
6 & 6 & 0 & 18 & 0.990 & 2.2e-3 & 853776 & 0.255 & 0.279\\

1 & 6 & 5 & 8  & 0.991 & 2.0e-3 &  11088 & 0.405 & 0.866\\
2 & 6 & 4 & 10 & 0.991 & 1.8e-3 &  60984 & 0.448 & 0.765\\
3 & 6 & 3 & 9  & 0.992 & 1.7e-3 & 203280 & 0.332 & 0.702\\
4 & 6 & 2 & 20 & 0.991 & 9.6e-4 & 457380 & 0.223 & 0.553\\
5 & 6 & 1 & 16 & 0.991 & 1.6e-3 & 731808 & 0.100 & 0.462\\
5 & 7 & 2 & 22 & 0.993 & 2.0e-3 & 627264 & 0.313 & 0.244\\
4 & 7 & 3 & 20 & 0.992 & 1.1e-3 & 392040 & 0.481 & 0.093\\
4 & 8 & 4 & 22 & 0.991 & 1.9e-3 & 245025 & 0.349 & 0.278\\

\addlinespace[2pt]
\midrule
\multicolumn{9}{c}{\textbf{$3\times 4$}}\\
\midrule
1 & 1 & 0 & 3  & 0.991 & 4.5e-3 &    144 & 0.739 & 0.969\\
2 & 2 & 0 & 5  & 0.991 & 2.4e-3 &   4356 & 0.189 & 0.899\\
3 & 3 & 0 & 6  & 0.991 & 2.4e-3 &  48400 & 0.578 & 0.823\\
4 & 4 & 0 & 20 & 0.990 & 9.3e-4 & 245025 & 0.038 & 0.584\\
5 & 5 & 0 & 23 & 0.990 & 7.3e-4 & 627264 & 0.184 & 0.553\\
6 & 6 & 0 & 19 & 0.992 & 1.9e-3 & 853776 & 0.252 & 0.272\\

1 & 6 & 5 & 6  & 0.992 & 2.1e-3 &  11088 & 0.404 & 0.879\\
2 & 6 & 4 & 7  & 0.991 & 2.4e-3 &  60984 & 0.244 & 0.785\\
3 & 6 & 3 & 10 & 0.991 & 1.5e-3 & 203280 & 0.366 & 0.686\\
4 & 6 & 2 & 12 & 0.992 & 1.7e-3 & 457380 & 0.082 & 0.589\\
5 & 6 & 1 & 30 & 0.990 & 4.8e-4 & 731808 & 0.146 & 0.477\\
5 & 7 & 2 & 22 & 0.991 & 1.9e-3 & 627264 & 0.487 & 0.269\\

\end{longtable}
\endgroup

% Naga Oka theorem

Several trends emerge from the sector-resolved data. The minimum circuit depth is not determined solely by the Hilbert-space dimension. While larger sectors generally require more evolution layers, sectors with comparable dimensions can exhibit significantly different variational complexity. The strongest correlations are observed with magnetization and the overlap between the interacting ground state and the non-interacting Slater determinant.

For one-dimensional systems, the easiest sectors are typically found at low particle numbers, particularly the $(1,1)$ sector, where the interacting ground state remains close to the non-interacting reference state and can therefore be represented accurately using relatively few HVA layers. In contrast, the most demanding sectors occur in highly asymmetric spin configurations containing a minority-spin hole embedded in a polarized background, such as $(1,L-2)$. These sectors exhibit small Slater overlaps and require substantially larger values of $S_{\min}$ to achieve the target fidelity.

The many-body gap and entanglement entropy provide additional indicators of complexity but do not individually determine the required circuit depth. Instead, the variational complexity appears to arise from a combination of correlation strength, magnetization, and the distance between the non-interacting and interacting ground states.

The two-dimensional results exhibit a different pattern of variational complexity than the one-dimensional chains. While highly polarized sectors are often the most demanding in one dimension, the largest values of $S_{\min}$ in the $2\times6$ and $3\times4$ lattices occur predominantly in sectors with less magnization or balanced number of spins near half filling. These sectors combine large Hilbert-space dimensions with strong correlation effects, resulting in smaller overlaps with the non-interacting Slater determinant and therefore requiring deeper variational circuits. For example, in the $2\times6$ lattice the $(5,5)$ sector requires substantially more evolution layers than dilute sectors such as $(1,1)$ despite both being represented within the same ansatz framework. The complete sector-resolved data provided here can therefore be used as a quantitative benchmark of HVA performance across the Hubbard-model Hilbert space.

%%%%%%%%%%%%%%%%%%%%%%%%%%%%%%%%%%%%%%%%%%%%%%%%%%%%%%%%%%%%%%%%%%%%%%%%%%%
\begin{figure*}
    \centering
    \includegraphics[width=0.9\linewidth]{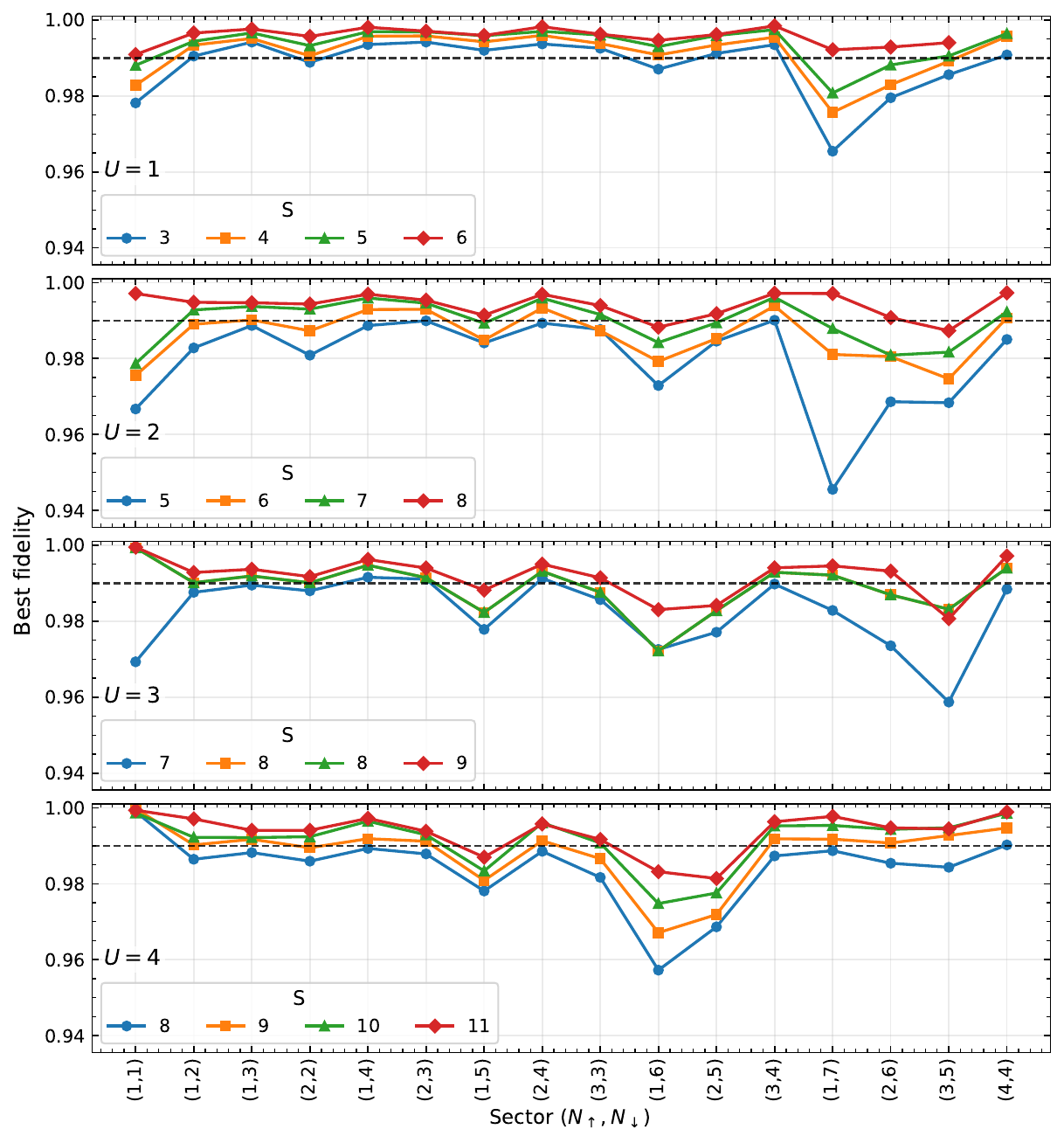}
\caption{Best variational fidelity obtained for each particle-number sector $(N_\uparrow,N_\downarrow)$ of the $L=8$ one-dimensional Hubbard chain as a function of the number of evolution layers $S$, for interaction strengths $U=1,2,3,$ and $4$. Each panel corresponds to a fixed value of $U$, while different curves represent different evolution depths $S$. For every sector and $S$ value, the highest fidelity obtained among all optimization runs is shown. The horizontal dashed line marks the target fidelity threshold ($\mathcal{F}=0.99$). As the interaction strength increases, larger evolution depths are generally required to achieve the same target fidelity, although the required depth varies significantly between particle-number sectors. Sectors near half filling and with small magnetization typically reach the target fidelity with fewer layers, whereas more challenging sectors require deeper circuits.}
\label{fig:sec_diff_1d}
\end{figure*}
%%%%%%%%%%%%%%%%%%%%%%%%%%%%%%%%%%%%%%%%%%%%%%%%%%%%%%%%%%%%%%%%%%%%%%%%%%%%

For one dimensional system, the important observation is not just that larger $U$ requires more layers. In the data, sectors such as $(1,6), (2,5), (1,7),$ and $(3,5)$ are systematically harder than neighboring sectors, whereas half-filled balanced sectors are often easier despite having much larger Hilbert-space dimensions.

%%%%%%%%%%%%%%%%%%%%%%%%%%%%%%%%%%%%%%%%%%%%%%%%%%%%%%%%%%%%%%%%%%%%%%%%%%%
\begin{figure*}
    \centering
\includegraphics[width=0.9\linewidth]{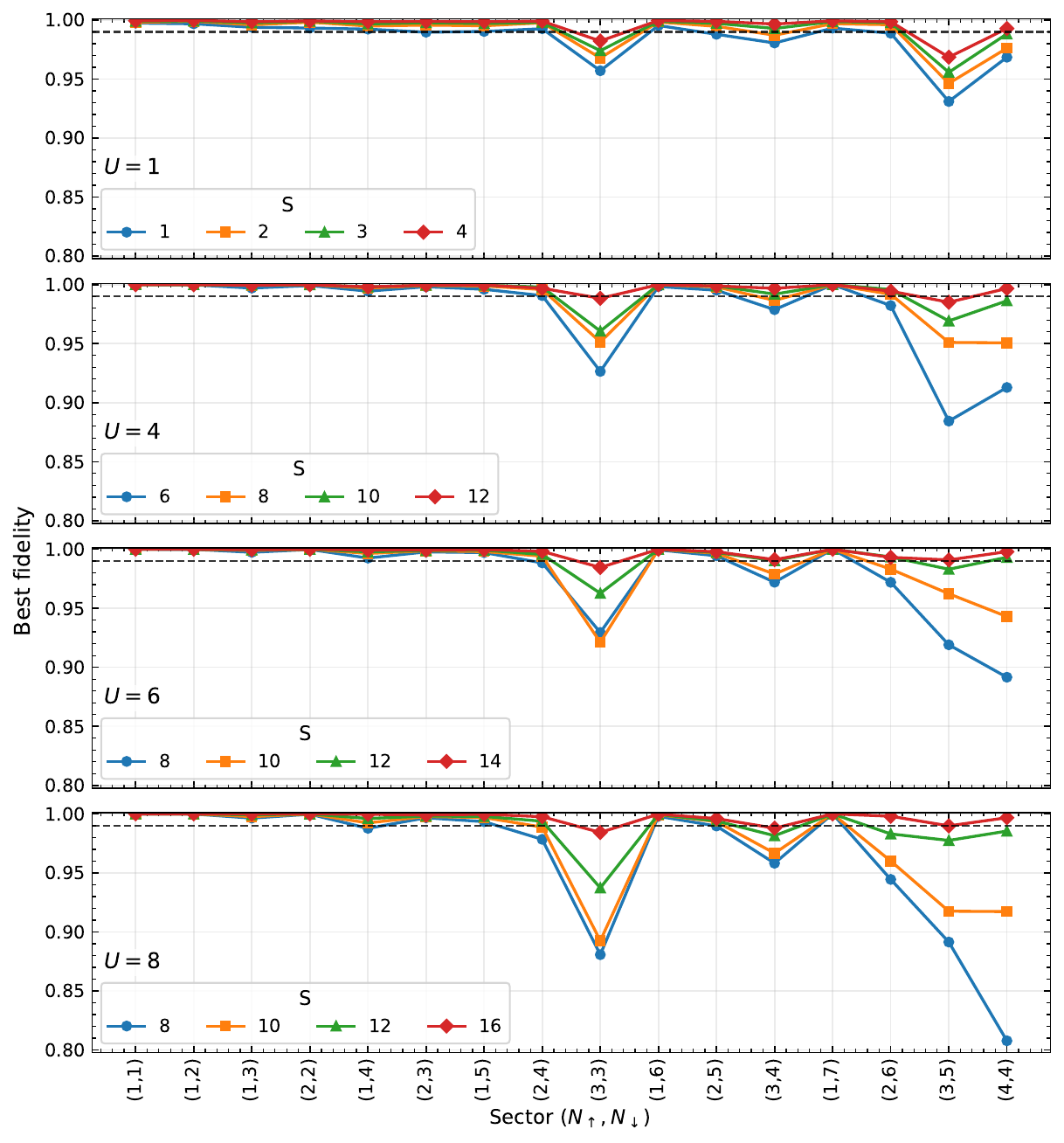}
\caption{
Best fidelity obtained in each symmetry-inequivalent sector of the $2\times4$ Hubbard lattice as a function of the number of HVA evolution layers $S$ for interaction strengths $U=1$, $4$, $6$, and $8$. The dashed line marks the target fidelity $\mathcal{F}=0.99$. Compared with the one-dimensional chain, the fidelity exhibits a stronger dependence on both interaction strength and spin sector. Near half filling, several sectors require substantially deeper circuits, reflecting the increased complexity of two-dimensional correlations and the longer JW strings associated with the fermion-to-qubit mapping.
}
\label{fig:sec_diff_2d}
\end{figure*}
%%%%%%%%%%%%%%%%%%%%%%%%%%%%%%%%%%%%%%%%%%%%%%%%%%%%%%%%%%%%%%%%%%%%%%%%%%%%

In contrast to one dimensional system, the striking feature here is that the hardest sectors are no longer the highly polarized sectors. Instead, sectors closer to half filling become the bottleneck. This is consistent with stronger correlation effects in two dimensions, where magnetic fluctuations and competing many-body configurations become more important.

Figures~\ref{fig:sec_diff_1d} and \ref{fig:sec_diff_2d} show the
best fidelity obtained in each symmetry-inequivalent spin sector as a
function of the number of HVA evolution layers for several interaction
strengths. In all cases, increasing the number of evolution layers
systematically improves the fidelity, but the rate of convergence
depends strongly on both the interaction strength and the spin sector.

A notable feature is that the sectors requiring the largest number of
evolution layers are not fixed as the interaction strength is varied.
At weak coupling, many sectors remain close to the non-interacting
Slater determinant and can therefore be represented accurately with
relatively shallow circuits. As $U$ increases, the character of the
ground state changes and different sectors become increasingly
correlated. Consequently, the sectors that are most difficult to
prepare at one interaction strength are not necessarily the most
difficult at another.

This behavior indicates that variational complexity is not determined
solely by the Hilbert-space dimension or the particle number of a
sector. Instead, it reflects the interaction-dependent structure of
the many-body ground state. Quantities such as the overlap with the
non-interacting Slater determinant, the magnetization, and the many-body
correlation strength provide more useful indicators of the required
circuit depth. In particular, sectors with small Slater overlap
typically require substantially more evolution layers to achieve the
target fidelity.

The effect is especially pronounced in the two-dimensional lattices,
where sectors near half filling become progressively more difficult as
the interaction strength increases. This trend is consistent with the
enhanced correlation effects expected in this regime and explains why
the hardest sectors observed at $U=4$ differ from those at smaller
interaction strengths.

For the one-dimensional chains, highly spin-imbalanced sectors
frequently become the most demanding at intermediate and strong
interactions. However, the ordering of sector difficulty changes with
$U$, demonstrating that no single sector can be regarded as universally
hardest across the entire interaction range. For the two-dimensional geometries, the most demanding sectors are generally found closer to half filling. As the interaction strength is increased, additional sectors enter the strongly correlated regime and require deeper circuits, leading to a more pronounced redistribution of sector complexity than observed in one dimension.

The complete sector-resolved diagnostics, including $S_{\min}$,
magnetization, many-body gap, entanglement entropy (1D only), and
Slater overlap, are reported in Tables~\ref{tab:sector_depth_gap_ent_1d}
and \ref{tab:sector_depth_gap_2d}. Together with
Figs.~\ref{fig:sec_diff_1d} and \ref{fig:sec_diff_2d}, these data show
that variational complexity depends on the interaction strength and
cannot be characterized by a single sector-independent scaling law.

\section{Geometry dependence of the XC functional}

%%%%%%%%%%%%%%%%%%%%%%%%%%%%%%%%%%%%%%%%%%%%%%%%%%%%%%%%%%%%%%%%%%%%%%%%%%%%%
\begin{figure*}
    \centering
\includegraphics[width=0.9\linewidth]{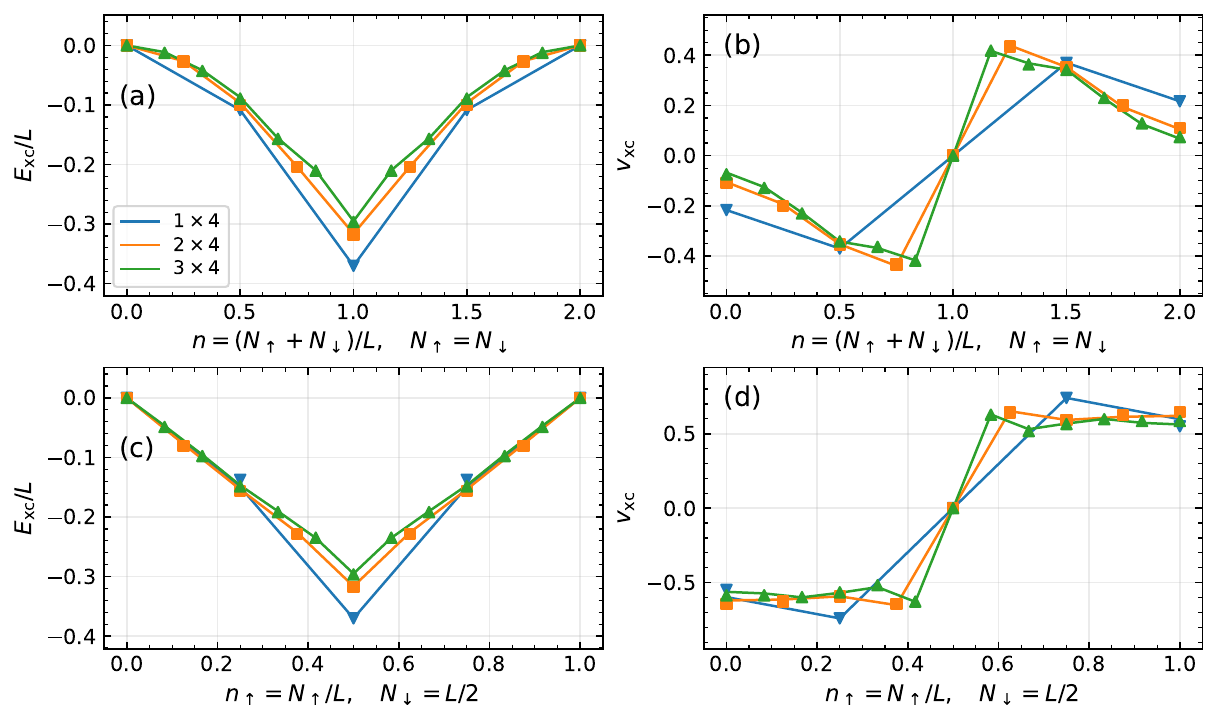}
\caption{
XC energy per site and XC potential per site obtained from exact diagonalization for the $1\times4$, $2\times4$, and $3\times4$ Hubbard lattices at $U=4$. Panels (a) and (b) show the nonmagnetic path $N_\uparrow=N_\downarrow$ as a function of total filling $n=(N_\uparrow+N_\downarrow)/L$. Panels (c) and (d) show the magnetic path obtained by fixing $N_\downarrow=L/2$ and varying $N_\uparrow$. The XC energy develops a pronounced minimum near half filling, while the XC potential exhibits a discontinuity associated with the Mott gap. Increasing the number of coupled chains reduces the magnitude of the XC energy, indicating the influence of lattice geometry and the gradual crossover from one-dimensional to two-dimensional correlation physics.
}
\label{fig:multi_chains}
\end{figure*}
%%%%%%%%%%%%%%%%%%%%%%%%%%%%%%%%%%%%%%%%%%%%%%%%%%%%%%%%%%%%%%%%%%%%%%%%%

Figure~\ref{fig:multi_chains} illustrates how the XC functional evolves as the lattice geometry changes from a one-dimensional chain to ladder and quasi-two-dimensional structures. The XC energy exhibits a pronounced minimum at half filling for all geometries, reflecting the enhanced correlation effects associated with the Mott-insulating regime. The depth of this minimum decreases systematically as additional chains are coupled, indicating a reduction in finite-size effects and a redistribution of correlation energy across the larger coordination environment.

The XC potential displays the characteristic jump near half filling expected from LDFT. While the jump remains visible for all geometries, its magnitude and shape depend on the lattice structure. The comparison demonstrates that the XC functional is strongly geometry dependent and therefore motivates the explicit construction of lattice-specific XC tables rather than relying on a universal parametrization.

%\twocolumngrid

\end{document}